\documentclass{aa} 
\usepackage{graphicx}
\usepackage{cancel}
\usepackage{txfonts}
\usepackage{natbib,twoopt}
\usepackage{ulem } %% to allow crossing out some text
\usepackage[breaklinks=true]{hyperref} %% to avoid \citeads line fills
\bibpunct{(}{)}{;}{a}{}{,}             %% natbib format for A&A and ApJ
\makeatletter
  \newcommandtwoopt{\citeads}[3][][]{\href{http://adsabs.harvard.edu/abs/#3}%
    {\def\hyper@linkstart##1##2{}%
     \let\hyper@linkend\@empty\citealp[#1][#2]{#3}}}
  \newcommandtwoopt{\citepads}[3][][]{\href{http://adsabs.harvard.edu/abs/#3}%
    {\def\hyper@linkstart##1##2{}%
     \let\hyper@linkend\@empty\citep[#1][#2]{#3}}}
  \newcommandtwoopt{\citetads}[3][][]{\href{http://adsabs.harvard.edu/abs/#3}%
    {\def\hyper@linkstart##1##2{}%
     \let\hyper@linkend\@empty\citet[#1][#2]{#3}}}
  \newcommandtwoopt{\citeyearads}[3][][]%
    {\href{http://adsabs.harvard.edu/abs/#3}
    {\def\hyper@linkstart##1##2{}%
     \let\hyper@linkend\@empty\citeyear[#1][#2]{#3}}}
\makeatother
\usepackage{xspace}
\usepackage{float}
\usepackage{color} 
\usepackage{ulem}
\usepackage{graphicx}	% Including figure file
\usepackage{msc}

\newcommand{\ie}{i.e.\xspace}
\newcommand{\eg}{e.g.\xspace}

\def\starlight{\textsc{starlight}\xspace}

\newcommand{\msun}{\ifmmode \text{M}_{\odot} \else M$_{\odot}$\fi\xspace}

\newcommand{\Mstar}{\ifmmode M_{\star} \else $M_{\star}$\fi\xspace}
\newcommand{\MBH}{\ifmmode M_{\mathrm{BH}} \else $M_{\mathrm{BH}}$\fi\xspace}
\newcommand{\mdot}{\ifmmode \dot{m} \else \dot{m}\fi\xspace}
\newcommand{\Lrad}{\ifmmode L_\mathrm{1.4} \else $L_\mathrm{1.4}$\fi\xspace}
\newcommand{\Frad}{\ifmmode F_\mathrm{1.4} \else $F_\mathrm{1.4}$\fi\xspace}
\newcommand{\LHa}{\ifmmode L_{\Ha} \else $L_{\Ha}$\fi\xspace}
\newcommand{\Loiii}{\ifmmode L_{\oiii} \else $L_{\oiii}$\fi\xspace}
\newcommand{\Lbol}{\ifmmode L_\mathrm{bol} \else $L_\mathrm{bol}$\fi\xspace}
\newcommand{\Lbolmod}{\ifmmode L_\mathrm{bol}^\mathrm{mod} \else $L_\mathrm{bol}^\mathrm{mod}$\fi\xspace}
\newcommand{\Ledd}{\ifmmode L_\mathrm{Edd} \else $L_\mathrm{Edd}$\fi\xspace}
\newcommand{\Lmech}{\ifmmode L_\mathrm{mech} \else $L_\mathrm{mech}$\fi\xspace} 
\newcommand{\Avneb}{\ifmmode A_{V}^\mathrm{neb} \else $A_{V}^\mathrm{neb}$\fi\xspace} 
\newcommand{\Avstar}{\ifmmode A_{V}^\star \else $A_{V}^\star$\fi\xspace} 
\newcommand{\sigstar}{\ifmmode \sigma_{\star} \else $\sigma_{\star}$\fi\xspace} 

\newcommand{\hii}{\ifmmode \mathrm{H}\,\textsc{ii} \else H~{\sc ii}\fi\xspace}
\newcommand{\heii}{\ifmmode \mathrm{He}\,\textsc{ii} \else He~{\sc ii}\fi\xspace}
\newcommand{\Ha}{\ifmmode \mathrm{H}\alpha \else H$\alpha$\fi\xspace}
\newcommand{\Hb}{\ifmmode \mathrm{H}\beta \else H$\beta$\fi\xspace}
\newcommand{\Heii}{\ifmmode \mathrm{He}\textsc{ii} \else He~{\sc ii}\fi\xspace}
\newcommand{\oiii}{\ifmmode [\mathrm{O}\,\textsc{iii}] \else [O~{\sc iii}]\fi\xspace}
\newcommand{\oii}{\ifmmode [\mathrm{O}\,\textsc{ii}] \else [O~{\sc ii}]\fi\xspace}
\newcommand{\oi}{\ifmmode [\mathrm{O}\,\textsc{i}] \else [O~{\sc i}]\fi\xspace}
\newcommand{\nii}{\ifmmode [\mathrm{N}\,\textsc{ii}] \else [N~{\sc ii}]\fi\xspace}
\newcommand{\sii}{\ifmmode [\mathrm{S}\,\textsc{ii}] \else [S~{\sc ii}]\fi\xspace}
\newcommand{\Oiii}{[O~{\sc iii}]$\lambda$5007\xspace}
\newcommand{\Nii}{[N~{\sc ii}]$\lambda$6584\xspace}

\newcommand{\woiii}{\ifmmode W[\mathrm{O}\,\textsc{iii}] \else $W[\mathrm{O}\,\textsc{iii}]$\fi\xspace}
\newcommand{\wha}{\ifmmode W(\mathrm{H}\alpha) \else $W(\mathrm{H}\alpha)$\fi\xspace}

\newcommand{\ania}[1]{{\color{teal} #1}}

\newcommand*{\genbf}[1]{\ifmmode\pmb{#1}\else\bfseries{#1}\fi}
\newcommand{\comment}[1]{{\genbf #1}}

\titlerunning{Distribution of radio-loudness parameter among AGNs in the local Universe}
\authorrunning{A. W\'ojtowicz et al.}

\begin{document}

   \title{Implications of the continuous radio-loudness distribution \\ among AGNs in the local Universe}
   
   \authorrunning{A. W\'ojtowicz et al.}

   \subtitle{}

\author{A. W\'ojtowicz
        \inst{\ref{brno}}
        \and
        N. Vale Asari
        \inst{\ref{ufsc}}
        \and
        \L{}. Stawarz
        \inst{\ref{krakow}}
        \and
        G. Stasi\'nska
        \inst{\ref{obspm}}
        \and
        D. Kozie\l-Wierzbowska
        \inst{\ref{krakow}}
      }
                
\institute{Department of Theoretical Physics and Astrophysics, Faculty of Science, Masaryk University, Kotláršká 2, Brno, 61137, Czechia\label{brno}\\
           \email{awojtowicz@oa.uj.edu.pl}
           \and
           Departamento de F\'{\i}sica--CFM, Universidade Federal de Santa Catarina, C.P.\ 5064, 88035-972, Florian\'opolis, SC, Brazil\label{ufsc}
           \and           
           Astronomical Observatory, Jagiellonian University, ul. Orla 171, PL-30244 Krak\'ow, Poland\label{krakow}
           \and
           LUX, Observatoire de Paris, Universit\'e PSL, Sorbonne Universit\'e, CNRS, 92190 Meudon, France\label{obspm}
           }
           
   \date{Received ; accepted }

% \abstract{}{}{}{}{} 
% 5 {} token are mandatory
 
  \abstract
   {}
   {}
   {}
   {}
   {}

\abstract
 {}
 %We analyze the radio loudness of a large, homogeneously selected sample of radio galaxies from the Radio sources associated with Optical Galaxies and having Unresolved or Extended morphologies (ROGUE) I and II catalogues, which include both compact and extended radio sources, regardless of the level of their AGN activity.}
{
We investigate the radio loudness ($\mathcal{R}$) distribution in a large, homogeneous sample of radio galaxies.}  %and its dependence on the accretion rate ($\lambda$) and black‑hole mass (\MBH).}
 %{\comment {I would take out all this until "we investigate...} \comment{Agreed} \ania{ We seek to answer the question of whether a bimodal distribution in $\mathcal{R}$ characteristic for quasar is also common among radio galaxies characterized by much lower accretion rates.} \comment{I don't think I agree - is the bimodality settled for quasars?} \ania{My impression was that majority of people believe it is there, and hence discuss radio-loud and radio-quiet sources.} We investigate the radio loudness ($\mathcal{R}$) distribution and its dependence on  observationally constrained parameters, such as the accretion rate ($\lambda$), black‑hole mass (\MBH) and galaxy host morphology, and interpret the observed relations and scatter in our data in the framework of theoretical models.}
 {%The sample is taken from the ROGUE I and II catalogues, which have been constructed by cross matching the Sloan Digital Sky Survey galaxy catalogue with the FIRST and NVSS. 
 The sample is composed of galaxies from the ROGUE I/II catalogue belonging to the SDSS MGS and is divided into optically inactive radio galaxies (OPIRGs), optically active ones (OPARGs) and 'radio Seyferts'. We use optical, mid-infrared  and radio data to calculate the AGN bolometric luminosities, accretion rate ($\lambda$), black‑hole mass (\MBH) and $\mathcal{R}$.}
%are defined via the $D_n(4000)$–$L_{1.4}/M_\star$ plane, H$\alpha$ equivalent widths and position in the BPT diagram.}
%  {Our radio galaxy sample is taken from the Radio sources associated with Optical Galaxies and having Unresolved or Extended morphologies (ROGUE) I and II catalogues, which have been constructed by cross matching the Sloan Digital Sky Survey galaxy catalogue with the First Images of the Radio Sky at Twenty cm observations and the NRAO VLA Sky Survey. We use optical, mid-infrared (from the Wide-field Infrared Survey Explorer) and radio data to calculate the AGN bolometric lumosities, $\mathcal{R}$, $\lambda$ and \MBH. Our subsamples of 'radio Seyferts', optically active and optically inactive radio galaxies (OPARGs and OPIRGs) are defined via the $D_n(4000)$–$L_{1.4}/M_\star$ plane, H$\alpha$ equivalent widths \comment{and position in the BPT diagram.}}
 {Contrary to \comment{some} previous studies based on restricted samples, using our complete sample of objects with redshifts $z < 0.4$, we find no evidence of bimodality in $\mathcal{R}$. The highest $\mathcal{R}$ values are  associated with extended radio structures. 
 %(FRI/FRII/hybrids). 
 $\mathcal{R}$ is anti‑correlated with $\lambda$, and spans $\sim 2$~dex at fixed $\lambda$. Radio Seyferts, OPARGs and OPIRGs form a sequence of increasing \MBH with substantial overlap. Radio Seyferts show no $\mathcal{R}$--\MBH correlation, whereas OPARGs and OPIRGs show a weak positive trend. From theoretical considerations, the observed $\sim 2$-dex spread in radio luminosity and $\mathcal{R}$ can be reproduced by only a $\sim 4$-fold variation in the dimensionless magnetic flux $\varphi$ assuming realistic black‑hole spins.}
% {Contrary to previous studies based on restricted samples, we find no evidence of bimodality in $\mathcal{R}$. The highest $\mathcal{R}$ values are almost exclusively associated with extended radio structures (FRI/FRII/hybrids). $\mathcal{R}$ is anti‑correlated with $\lambda$, and spans $\sim 2$--3 dex at fixed $\lambda$, with scatter increasing at lower $\lambda$ and higher $\mathcal{R}$ values. Spiral and lenticular  galaxies occupy the high‑$\lambda$, low‑$\mathcal{R}$ region, while ellipticals reside at lower $\lambda$, higher $\mathcal{R}$ and show larger dispersion. Radio Seyferts, OPARGs and OPIRGs form a sequence of increasing \MBH with substantial overlap. Radio Seyferts show no $\mathcal{R}$-\MBH correlation, whereas OPARGs and OPIRGs show a weak positive trend. From scaling relations derived from theoretical considerations, the observed $\sim 2$-dex spread in radio luminosity and $\mathcal{R}$ can be reproduced by only a $\sim 4$-fold variation in the dimensionless magnetic flux $\varphi$, assuming `realistic' black‑hole spins $a\approx0.6$–1.0.}
 {The smooth distribution of radio loudness supports a common evolutionary path for all radio sources, with black hole spin and magnetic field varying continuously. The radio loudness depends on black hole mass and accretion rate, while moderate variations in $\varphi$ may account for the observed scatter in this relation.}
 % {The smooth distribution of radio loudness strongly supports the notion of a common evolutionary path for all types of radio sources, with black hole spin and magnetic field varying continuously across the population. The radio loudness depends on black hole mass and accretion rate, while moderate variations in $\varphi$ may account for the observed scatter in this relation.}


   \maketitle
%
%-------------------------------------------------------------------

\section{Introduction}

%\ania{Ania}
%\dorota{Dorota}
%\comment{Grazyna}
%\lukasz{Lukasz}
%\comment{Natalia}

Active galactic nuclei (AGN) span a wide range of accretion rates, reflected in the large spread of optical emission from their accretion flows and disks. They also show significant differences in radio emission, associated with relativistic jets and outflows launched by the central engine \citep[see \eg][for a review]{Tadhunter16,Hardcastle2020,Saika2022}. Notably, even among sources with comparable accretion rates, the efficiency of radio production can differ by several orders of magnitude.

To quantify this efficiency, \citet{Kellermann.etal.1989a} introduced the radio-loudness parameter denoted here $\cal{R}_\mathrm{K}$, defined as the ratio of 5~GHz radio to optical $B$-band flux density. This parameter has since been widely applied to different AGN classes \citep[\eg][]{Sikora.Stawarz.Lasota.2007a, Rafter.etal.2009a, Singh2018, Rusinek.etal.2020a} as a proxy for jet production efficiency, under the assumption that radio emission traces jets and that optical light is dominated by the accretion disk. These assumptions are generally valid for luminous quasars, but not always for lower-luminosity AGN and quasars.
For example, in Seyfert galaxies, particularly those accreting at sub-Eddington rates, the host galaxy can contribute significantly to the observed optical flux, alongside the AGN emission \citep[see \eg][]{Ho01}. Likewise, star formation processes may contribute to, or even dominate, the radio emission rather than jets. Even in quasars with very low radio luminosities, star formation may outshine jet-related radio output \citep{Kellermann16,Rankine2021}. 
The original definition of the radio-loudness parameter may thus not be directly applicable in all cases and warrants careful consideration.

The distribution of radio-loudness has been the subject of debate for decades. \citet{Kellermann.etal.1989a} argued for a bimodal distribution in quasars, with radio-loud and radio-quiet populations separated at $\mathcal{R}_\mathrm{K} = 10$. Later studies revisited this claim with mixed results, some supporting a dichotomy \citep[see \eg][]{Ivezic02, Sikora.Stawarz.Lasota.2007a, Rafter.etal.2009a, Rusinek.etal.2020a} and others attributing the apparent bimodality to selection effects and flux limits \citep[see \eg][and references therein]{Mahony12, Singal.etal.2013a, Macfarlane2021}.

While the question of radio-loudness bimodality remains unsettled for quasars, it is clear that their radio-loudness distribution is broad, spanning several orders of magnitude.
The radio-loudness parameter has been extensively studied in the literature for all types of AGNs \citep[e.g.][]{Miller.etal.1990a, White.etal.2000a, White.etal.2007a, Sikora.Stawarz.Lasota.2007a, Rafter.etal.2009a, Singal.etal.2013a, Rusinek.etal.2020a}, with entirely different conclusions about its distribution. It was suggested that radio-loud and radio-quiet sources  may have different evolutionary paths or different jet production mechanisms \citep{Bicknell.2002a}.

Much less is known about radio-loudness in AGNs accreting below the Eddington limit, such as Seyfert galaxies and low luminosity radio galaxies. Several works have pointed out systematic differences: Seyferts, typically found in spiral and lenticular hosts, never reach the high radio-loudness values observed in radio galaxies, which are usually associated with ellipticals \citep{Xu99, Sikora.Stawarz.Lasota.2007a,Koziel2017,Zheng2020}.
Moreover, Seyferts and radio galaxies follow a similar overall anti-correlation between $\mathcal{R}_\mathrm{K}$ and Eddington ratio $\lambda$ (\ie the ratio of total radiative output to the Eddington luminosity for the accreting black hole), but the separation in the $\mathcal{R}_\mathrm{K}$–$\lambda$ plane between them cannot be explained solely by selection effects. If very radio-loud Seyferts with $\mathcal{R}_\mathrm{K} > 10^4$ existed, they should be detectable, yet none are observed.

Progress therefore requires large and homogeneous samples of non-quasar AGNs, with careful treatment of host-galaxy contamination in both the radio and optical bands, to establish the true shape of the radio-loudness distribution and its dependence on accretion rate and host properties.
\comment{Unlike previous works, }we present a study of the radio-loudness 
%in a sample of galaxies from the Main Galaxy Sample  of the Sloan Digital Sky Survey (SDSS; \citealp{York.etal.2000a}) which appear in the ROGUE I and II catalogues of radio sources (\citealp{KozielWierzbowska.Goyal.Zywucka.2020a}; 2025, \comment{to be submitted}.). 
in  galaxies in the local Universe \comment{with low to moderate accretion rates}, selected from the the ROGUE I and II catalogues of radio sources (\citealp{KozielWierzbowska.Goyal.Zywucka.2020a}; 2025, to be submitted) that belong to the Main Galaxy Sample  of the Sloan Digital Sky Survey (SDSS; \citealp{York.etal.2000a}). 
Our sample is homogeneously selected with both optical and radio flux limits well defined, and contains Seyfert galaxies as well as optically active and optically inactive radio galaxies (OPARGs and OPIRGs), whose redshifts ($z$) are in the $0.002 \leq z \leq 0.4$ range. The sample and its selection effects are described in Section~\ref{sec:Data}). \comment{Construction of this catalog results in exclusion of Broad Line Radio Galaxies.} Such a sample allows us to study the radio-loudness distribution and its implications for the evolution of extragalactic radio sources in the current epoch (see Section~\ref{sec:Discussion}).

Throughout this work, we adopt a cosmology with
$H_0 = 70 \ {\rm km} \ {\rm s}^{-1} \ {\rm Mpc}^{-1}$, $\Omega_M = 0.30$, and $\Omega_{\Lambda} = 0.70$.

\section{Data}
\label{sec:Data}

%------------------------------- Figure -------------------------------%
\begin{figure*}[!ht]
  \centering
  % [trim={left bottom right top},clip]
  \includegraphics[width=0.85\textwidth]{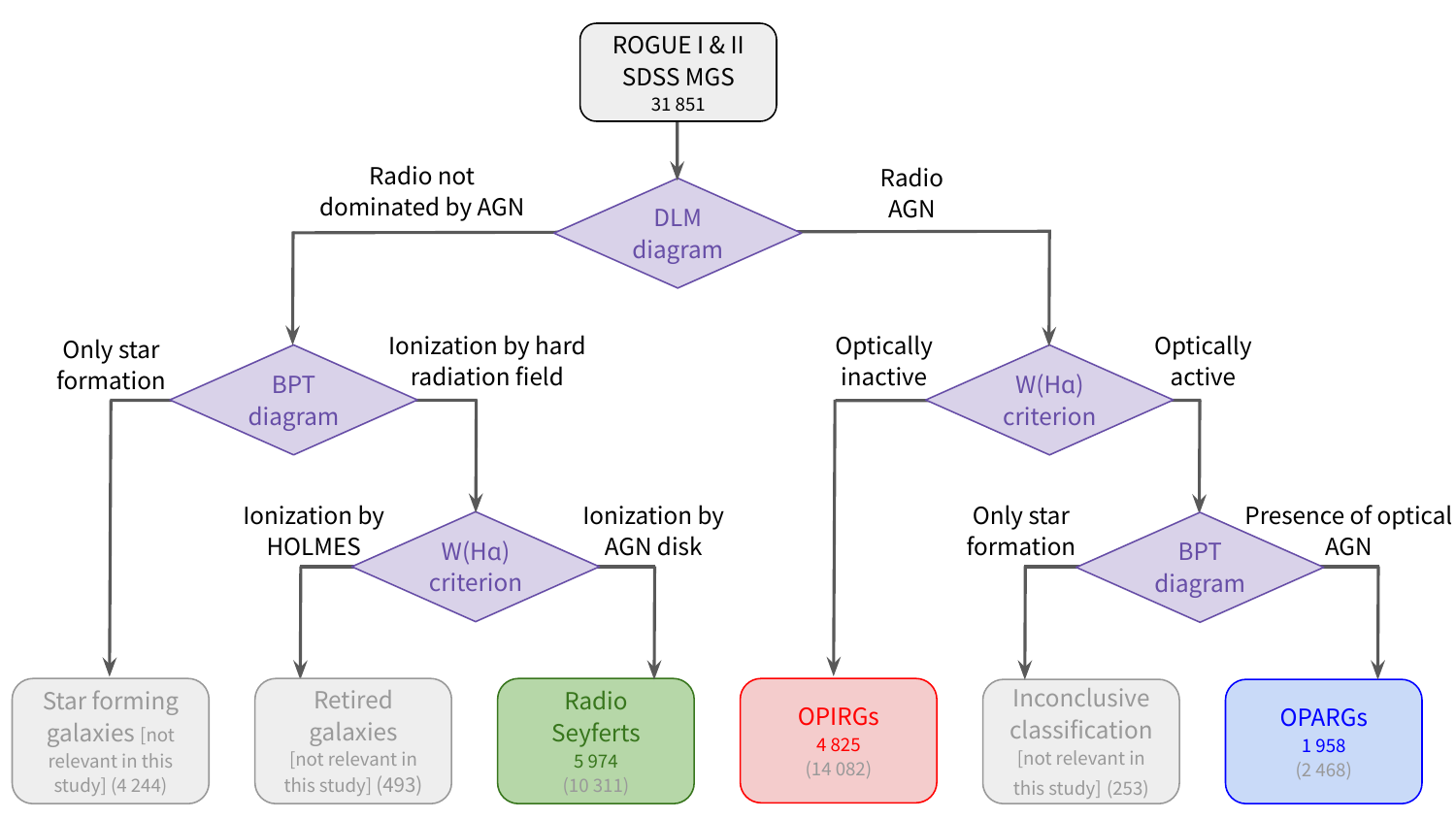}
  \caption{Flowchart detailing our sample selection. Retired galaxies are galaxies that have stopped forming stars and are ionized by their populations of hot low-mass evolved stars (HOLMES), see \citet{Stasinska.etal.2008a}. Similarly to the star-forming galaxies, they are not relevant for the present study. The parent `ROGUE I \& II SDSS MGS' sample already incorporates redshift range limits, optical spectral quality cuts, and removal of sources whose radio morphology is blended or non-detected.
  The number of sources in each subsample is given in the panels. Numbers in brackets show number counts before applying limits to the $S/N$ of \oiii or $W3$ luminosities, to emission line detection and to the star-formation-corrected radio luminosity values.}
  
  \label{fig:sample}
\end{figure*}
%------------------------------- Figure -------------------------------%

\subsection{Sample selection}
\label{sec:sample}

\subsubsection{Datasets and parent sample}

% 1. SDSS
We briefly summarize the optical, mid-infrared (MIR) and radio datasets used in this work, as well as our parent sample, which is the same as the one by \citet{Stasinska.etal.2025a}.
We select galaxies from the Main Galaxy Sample (MGS; \citealp{Strauss.etal.2002a})\footnote{Note that the MGS does not contain type I AGNs, which are in the quasar sample of the SDSS.} of the SDSS Data Release 7 (DR7; \citealp{Abazajian.etal.2009a}). 
We retain only galaxies whose redshifts ($z$) are in the $0.002 \leq z \leq 0.4$ range, to avoid luminosity distances being dominated by peculiar motions \citep{Ekholm.etal.2001a}, and to ensure that the \Ha emission line falls into the SDSS spectral range. The  \Ha  line is essential for the classification of  radio galaxies into optically active (OPARG) or optically inactive (OPIRG), as explained by \citet{Stasinska.etal.2025a}. We also select only galaxies with a signal-to-noise ratio $\geq 10$ in the continuum at 4020~\AA, to be able to safely derive their properties from their spectra.

Optical spectral data are taken from the \starlight\ database \citep{CidFernandes.etal.2005a}. 
Black hole masses (\MBH) are derived from stellar velocity dispersions measured by \starlight\ using the \citet{Tremaine.etal.2002a} relation. Stellar masses ($M_\star$) are obtained via \starlight's inverse population synthesis applied to restframed SDSS fibre spectra, then corrected to total galaxy values using $z$-band photometry. Star formation rates (SFR) over the past 100 Myr are estimated by summing the stellar masses of populations younger than this age and dividing by the time span, following \citet[eq.~10]{Asari.etal.2007a}. We exclude a small fraction ($< 0.3\%$) of galaxies with $M_\star \leq 10^7\, \mathrm{M_\odot}$ or non-positive values of `petroR50', the 50\,per\,cent Petrosian light radii in the $r$-band as available from the SDSS DR7 database. Our `MGS' parent sample  after the redshift and quality cuts is comprised of 616,069 objects.

% 2. WISE
MIR data are obtained by matching sources in the Wide-field Infrared Survey Explorer (WISE; \citealp{Wright.etal.2010a}) and SDSS DR7 galaxies within 1 arcsecond. The 12~$\mu$m luminosities ($W3$) are calculated assuming a constant power-law spectra (i.e. $F_\nu \sim \nu^0 = \text{constant}$) as in \citet[section 2.3]{KozielWierzbowska.etal.2021a}.

% 3. ROGUE
Radio data for our targets are taken from the ROGUE I and II catalogues (\citealp{KozielWierzbowska.Goyal.Zywucka.2020a}; 2025, to be submitted), which are based on SDSS DR7 galaxies with the continuum signal-to-noise ratio cut of 10. Radio counterparts were searched in the  First Images of the Radio Sky at Twenty cm \citep[FIRST;][]{White.etal.1997a} and the NRAO VLA Sky Survey \citep[NVSS;][]{Condon.etal.1998a} radio catalogues. ROGUE I contains those targets for which FIRST detects a compact core, while ROGUE II includes systems whose radio counterparts do not possess a core-like structure. Both catalogues provide 1.4~GHz radio fluxes
%luminosities ($L_{1.4}^\mathrm{total}$) 
and visual classifications of radio and optical morphologies. 
Total monochromatic 1.4~GHz radio luminosities ($L_{1.4}^\mathrm{total}$) are calculated assuming a spectral index of 0.75 as in \citet[equation 1]{KozielWierzbowska.etal.2021a}.
Sources whose radio morphology from ROGUE I \& II is classified as B (`blended') or ND (`not detected') are excluded. Our ROGUE I \& II SDSS MGS parent sample is comprised of 31,851 galaxies.

\subsubsection{Subsamples}

%------------------------------- Figure -------------------------------%
\begin{figure*}[!ht]
  \centering
  % [trim={left bottom right top},clip]
  \includegraphics[width=0.8\textwidth]{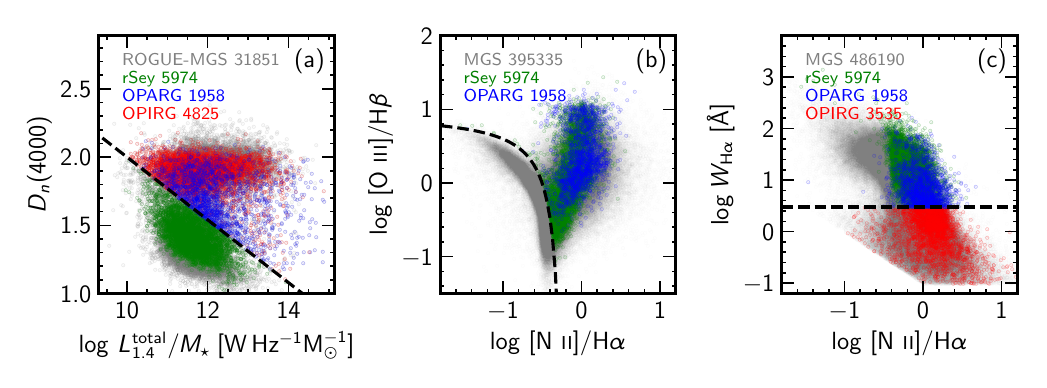}
  \caption{
    Diagrams showing the loci of our radio Seyferts (green), OPARGs (blue) and OPIRGs (red) subsamples in relation to our parent
    samples: ROGUE I \& II SDSS MGS (`ROGUE-MGS') for panel (a), SDSS MGS (`MGS') for panels (b) and (c) (grey points).
  (a) $D_n(4000)$ versus $L_{1.4}/M_\star$ (DLM) plane. The dashed line by \citet{KozielWierzbowska.etal.2021a} separates radio emission dominated (top) or not (bottom) by AGN.
  (b) \Oiii/\Hb versus \Nii/\Ha (BPT) plane. The dashed line by \citet{Stasinska.etal.2006a} separates galaxies ionized only by star formation (left) from those ionized by a hard ionizing radiation field (right). OPIRGs are not shown since they have not been selected based on this diagram.
  (c) \wha versus \Nii/\Ha (WHAN) plane. The dashed line shows the $\wha = 3$~\AA\ criterion by \citet{CidFernandes.etal.2011a} to separate ionization only by HOLMES (bottom) from ionization by other sources (top). OPIRGs lacking \nii detection are not shown on this plane. 
  }
  \label{fig:subsamples}
\end{figure*}
%------------------------------- Figure -------------------------------%

Following \citet{Stasinska.etal.2025a}, we define sub-samples of 2,468 OPARGs and 14,082 
OPIRGs based on a 3\,\AA\ threshold in \Ha equivalent width, \wha.  Both types are located on the `radio-AGN' side in %These are all selected as `radio-AGN' in 
the $D_n(4000)$ versus $L_{1.4}^\mathrm{total}/M_\star$ (DLM) diagram \citep{KozielWierzbowska.etal.2021a}, where $D_n(4000)$ is the 4000~\AA\ break from synthetic spectra \citep[e.g.,][]{Stasinska.etal.2006a} and $L_{1.4}^\mathrm{total}/$\Mstar is the quotient between the total luminosity at 1.4\,GHz and the stellar mass obtained by \starlight. Additionally OPARGs must lie to the right of the \citet{Stasinska.etal.2006a} division in the \Nii/\Ha\ versus \Oiii/\Hb (BPT; \citealp{Baldwin.Phillips.Terlevich.1981a}) diagram\footnote{Throughout this work, all emission lines are given in restframe air wavelengths.} , \ie they must be classified as galaxies hosting an optical AGN.  For OPIRGs, if they show emission lines, the  $\wha < 3\,\AA\ $  criterion indicates that the emission is not due to an AGN, but to ionization by hot low-mass evolved stars (HOLMES) present in the galaxy \citep[see][]{Stasinska.etal.2008a, Stasinska.etal.2025a}.

We also consider a third subsample of 10,311  dubbed `radio Seyfert galaxies' constituted by sources in the ROGUE I and II catalogues lying in the `star forming' region of the DLM diagram, but are located to the right of the \citet{Stasinska.etal.2006a} line on the BPT, and with $\wha > 3$\,\AA. The flowchart in Figure~\ref{fig:sample} shows the criteria used to define our subsamples, while Figure~\ref{fig:subsamples} shows our subsamples on the DLM, BPT and \wha versus \nii/\Ha (WHAN) planes.

\subsubsection{Derived data}
\label{sec:derived-data}

In the following we describe the calculated quantities used in this work. 
We fit an empirical linear relation between $\log L_{1.4}^\mathrm{total}$ and $\log \mathrm{SFR}$ for star-forming (SF) galaxies (\ie those classified as SF in both the DLM and BPT diagrams) following \citet{Murphy.etal.2011a}. This is used to subtract the contribution of star formation to the radio emission, yielding star-formation-corrected $L_{1.4}$ values for all sources. 
The fractional star formation contribution $f$ to $L_{1.4}^\mathrm{total}$ is only substantial for radio Seyferts, with a median value of $f = 0.36$ and interquartile range of $0.18$--$0.59$. This allows us to analyze radio Seyferts alongside radio galaxies where $L_{1.4}^\mathrm{total}$ is dominated by AGN.
The correction is around the order of 1\% for OPARGs and OPIRGs.
%, with a median value $f = 0.012$ (interquartile range $0.002$--$0.041$) for OPARG and $f = 0.008$ (interquartile range $0.001$--$0.032$) for OPIRG sources.}
Hereafter $L_{1.4}$ is used to represent the star-formation-corrected values.

For OPARGs and radio Seyfert galaxies, bolometric luminosities (\Lbol) are derived from \oiii\ luminosities using the \citet[eq.\ 2]{Stasinska.etal.2025a} relation. This relation results from a series of
photoionization models whose ionizing sources have spectral energy distributions (SEDs) based on the observed continua of unobscured Type I AGNs in the optical and X-rays \citep{Jin.Ward.Done.2012a, Jin.etal.2018b}.  In addition, we require a $S/N \geq 3$ in \oiii fluxes.

For OPIRGs,  we use the type 2 AGN relation  between the nuclear 12 $\mu$  luminosity  and \Lbol\ by \citet[Table 5]{Spinoglio.FernandezOntiveros.Malkan.2024a}. Since the nuclear 12 $\mu$  luminosity is not available for most of our objects we use the WISE  $W3$ luminosity as a surrogate, keeping in mind that the WISE measurements are affected by the contribution of the underlying galaxy. The latter, however, is expected to be small since OPIRGs experience little -- if any -- star formation. Only objects with $S/N \geq 3$ in $W3$ are kept. 

Eddington ratios are computed as $\lambda = \Lbol / \Ledd$, with $\Ledd = 1.3 \times 10^{31} \, (M_{\mathrm{BH}}/M_{\odot})$\,W. In this paper, radio loudness is defined as the ratio of the radio 1.4\,GHz monochromatic luminosity to the bolometric AGN disk luminosity, $\mathcal{R} = \nu_{1.4} L_{1.4} / \Lbol$. The \citet{Kellermann.etal.1989a} definition of the radio-loudness is therefore approximately $\mathcal{R}_\mathrm{K} \simeq 10^{6} \, \mathcal{R}$, assuming that the average spectral index of the radio continua  is $\alpha = 0.8$, and that the $B$-band luminosity constitutes roughly $10\%$ of the bolometric luminosity of the AGN disk. Hence the radio-loud threshold of $\mathcal{R}^{\mathrm{K}} = 10$ translates to $\log \mathcal{R} = -5$ in our convention.

For clarity, when we use \Lbol estimated from \oiii, which applies to the OPARG and radio Seyfert subsamples, we use the canonic symbols $\lambda$ and $\mathcal{R}$. When we mix \Lbol estimated either from \oiii or $W3$ (for OPIRGS), we denote those parameters $\Lbol^\prime$, $\lambda^\prime$ and $\mathcal{R}^\prime$ to indicate they are based on a heterogeneous set of observational parameters. Note that we compare \Lbol for OPARGS using both \oiii and $W3$ and find a good agreement, so the values are internally consistent.

In the remaining of the paper, we show only objects for which both $\Lbol^\prime$ $\lambda^\prime$  and $\mathcal{R}^\prime$ can be calculated after the $S/N$ cuts in either \oiii or $W3$ luminosities. We exclude sources for which the star-formation-corrected radio luminosities values are below the lowest $L_{1.4}^\mathrm{total}$ value in the `ROGUE-MGS' sample, \ie those with $L_{1.4} < 10^{20} \mathrm{\,W\,Hz^{-1}}$. Radio Seyferts and OPARGs are also required to have all BPT emission lines (\Oiii, \Hb, \Nii, \Ha) detected. 

Our final subsamples are thus comprised of 5,974 radio Seyferts, 4,825 OPIRGs and 1,958 OPARGs, as shown in Figure~\ref{fig:sample} and Figure~\ref{fig:subsamples}(a).

%\subsection{Selection effects}
\subsection{Impact of flux limits on subsamples}
\label{sec:fluxlimit}

%------------------------------- Figure -------------------------------%
\begin{figure*}[!ht]
  \centering
  % [trim={left bottom right top},clip]
  \includegraphics[width=\textwidth, trim=50 230 140 20, clip]{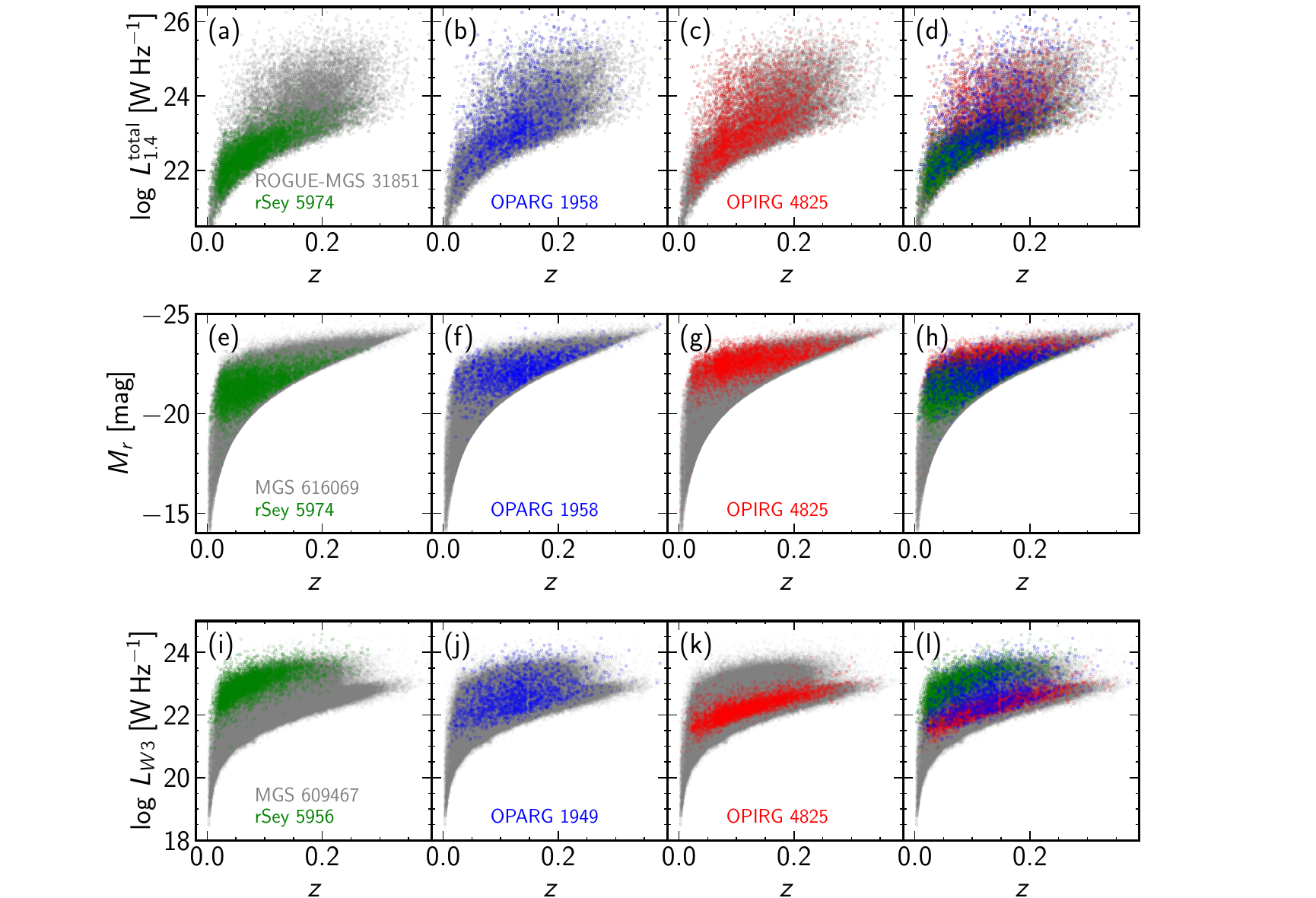}
  \caption{
  (a)--(d) Total radio luminosity,
  (e)--(h) absolute magnitude in the $r$-band, and 
  (i)--(l) mid-infrared $W3$ luminosity as a function of redshift. 
 Colours indicate the samples: radio Seyferts are in green, OPARGs in blue, and OPIRGs in red.
  The grey points indicate the parent samples: ROGUE I \& II SDSS MGS (`ROGUE-MGS') for the first row, SDSS MGS (`MGS') for the second and third rows. Numbers on the panel show the count of objects. Subsamples are shown overplotted on the parent sample on the first three columns; parent samples plus all subsamples are show together on the last column for comparison.}
  \label{fig:L-z}
\end{figure*}
%------------------------------- Figure -------------------------------%

Our ROGUE-MGS parent sample is subject to both  optical and radio flux limits.
%As our catalogue is based on the SDSS MGS catalogue and includes sources detected in radio by the NVSS survey, it is affected by flux limits in the optical and in radio. 
The SDSS MGS consists of galaxies with $r$-band Petrosian magnitudes $\leq 17.77$ and $r$-band Petrosian half-light surface brightnesses $\leq 24.5 \mathrm{\,mag\, arcsec^{-2}}$, with the latter cut removing only $0.1\%$ of the objects \citep{Strauss.etal.2002a}.
\citet{Strauss.etal.2002a} show that  overall the MGS is $\sim 95\%$ complete.
%{$> 99\%$ complete, with incompleteness arising primarily from confusion with nearby bright stars which tends to affect the brightest systems.}
Radio observations at 1.4~GHz have a flux limit of 1\,mJy for point sources (with synthesized beams of 5.''4 for FIRST and 45'' for NVSS; see \citealt{KozielWierzbowska.Goyal.Zywucka.2020a}). 

Figure~\ref{fig:L-z} shows $L_{1.4}^\mathrm{total}$, the absolute optical $r$-band magnitude ($M_r$) and the mid-infrared $W3$ luminosity ($L_{W3}$) versus redshift for our parent samples and subsamples.
%To better understand the impact of flux limits on our data, we examined the distribution of sources from the SDSS MGS and ROGUE catalogues (\ie those subject to the radio flux limit) in the $L_{1.4}^\mathrm{total}$, absolute $r$-band magnitude ($M_r$) and $W3$ luminosity ($L_{W3}$) versus redshift plane.
% We had removed this!
%, and in the $L_{1.4}^\mathrm{total}$ versus $M_r$ plane 
%\comment{in Figure~\ref{fig:L-z}.}
%The 1.4~GHz flux limit impacts the population of radio Seyfert galaxies in a different way than OPARGs and OPIRGs. Looking at panels (a) to (d) of Figure~\ref{fig:L-z}, we note Seyferts do not reach very high radio luminosities.
Panels (a) to (d) of Figure~\ref{fig:L-z} make it evident that radio
%It is evident that 
Seyfert galaxies are most significantly affected by the radio flux limit, likely with some fraction of the source population being undetected, as seen from the concentration of these sources near the lower edge of the ROGUE-MGS points. The population of OPARGs and OPIRGs are much less severely affected. 
We also note that the ROGUE-MGS parent sample and all subsamples show the same lower envelope of radio luminosity with redshift due to flux limits, indicating no faint-end bias.

Panels (e) to (h) of Figure~\ref{fig:L-z} show $M_r$ vs.\ $z$. 
The lower envelope for the parent sample is given by the SDSS optical flux limit, while the extra $M_r$ cut in the subsamples is a consequence of the radio flux limit. This effectively excludes r-band faint sources, whose emission is more likely dominated by star formation rather than AGN activity \citep{Capetti2022,Wojtowicz2023}.
%The lower envelope, common to the parent sample and all subsamples, is again a consequence of the optical flux limit. 
We also see that OPIRGs are dominated by the brightest galaxies in the sample at all redshifts; Seyfert galaxies are comprised of the faintest galaxies in the sample of radio detected MGS sources; and OPARGs are in between OPIRGs and Seyfert sources. 
Optically faint galaxies (greater $M_r$ values) drop out at higher redshifts due to the flux limit, while the brightest galaxies ($M_r \sim -24$), though included, are intrinsically rare due to the steep decline of the luminosity function at high luminosities \citep{Lumfuncellipticals,Ermash2013}.
%[*** this needs to be shortened: As expected, objects with low optical luminosity (greater $M_r$ values) rapidly drop out of MGS as the redshift increases, since extremely $r$-band bright galaxies are uncommon in most of the local Universe. Although the population of optically (in the $r$-band) faint galaxies is underrepresented across the considered redshift range, bright sources with absolute magnitudes down to $M_r \sim -24$ are still included in our survey. The brightest galaxies, although underrepresented at lower redshifts, are intrinsically rare, as the luminosity function declines steeply with increasing optical brightness \citep{Lumfuncellipticals,Ermash2013}.]

Panels (i) to (l) of Figure~\ref{fig:L-z} show $L_{W3}$ as a function of $z$. Some radio Seyferts and OPARGs are absent in those panels due to missing $W3$ data. $W3$ is however used to calculate $\Lbol$ only for OPIRGs, which lie away from both the lower envelope and the bright cloud of the parent MGS distribution. 

%Finally, we conclude that, although low radio-luminosity objects may be underrepresented, the radio-detected sources in our sample satisfactorily reflect the true distribution at the high radio-luminosity end. 

Although low radio-luminosity objects may be underrepresented, the radio-detected sources in our sample provide good coverage at the high radio-luminosity end. This regime matches the radio-loudness range investigated in previous studies and is most relevant for examining the radio-loudness distribution. The radio luminosity function is expected to decline at the faint end ($\log L_{1.4} < 22\,\mathrm{W\,Hz^{-1}}$), meaning that our sample effectively captures the bulk of the population.

\section{Discussion}
\label{sec:Discussion}

\subsection{There is no bimodality in the radio-loudness distribution}
%------------------------------- Figure -------------------------------%
\begin{figure*}[!ht]
  \centering
  % [trim={left bottom right top},clip]
  \includegraphics[width=0.7\textwidth, trim=0 10 0 0]{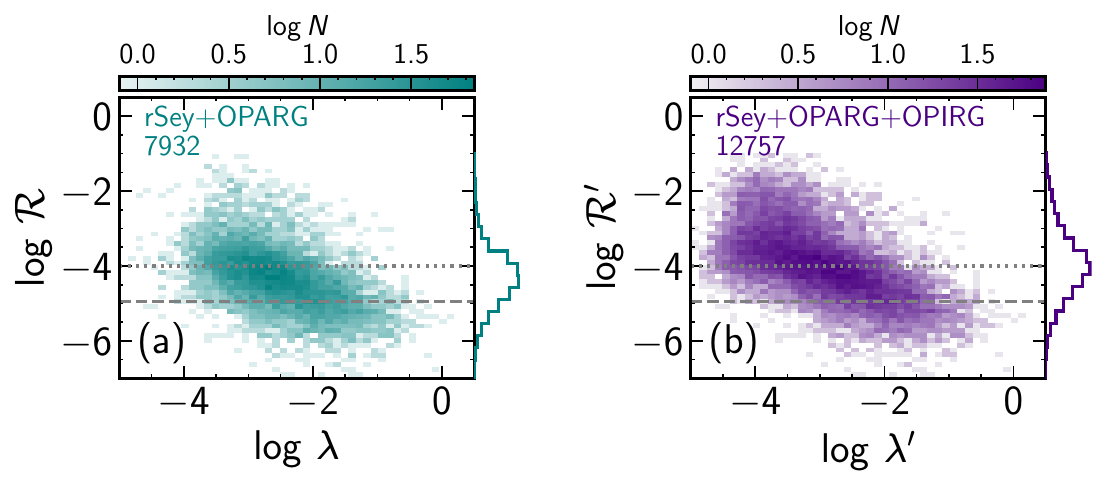}
  \caption{Two-dimensional histograms of radio loudness ($\mathcal{R}$) versus Eddington ratio ($\lambda$) for (a) radio Seyfert and OPARG objects, and (b) radio Seyfert, OPARG, and OPIRG objects combined. Bin colours represent the logarithm of the number of objects, as indicated by the colour bars. Marginal 1D histograms of radio loudness are shown on the right hand side of each panel. 
  Values with and without a prime indicate \Lbol\ derived solely from \oiii\ or from a mix of \oiii\ (for OPARG and Seyfert) and $W3$ (for OPIRG) luminosities, respectively. The number of objects is indicated in each panel. The dashed line corresponds to the classical radio-loudness threshold of \citet{Kellermann.etal.1989a}, and the dotted line to the division from \citet{Gupta2018}; both are shown using our convention.
  No bimodality is seen in either the 1D or 2D histograms.}
  \label{fig:Rlambda2D}
\end{figure*}
% ------------------------------- Figure -------------------------------%

Figure \ref{fig:Rlambda2D} shows 2D histograms of $\log {\cal R}$ vs $\log \lambda$, together with 1D histograms for radio-loudness. 
Panel (a) shows the radio-loudness and the Eddington ratio only for OPARG and radio Seyfert objects, for which bolometric luminosities are based on photoionization models with AGN SEDs (see Sect.~\ref{sec:derived-data}). Panel (b) includes OPIRGS, for which bolometric luminosities are inferred from $W3$ luminosities. The 2D histograms are proportional to the logarithm of number of objects in each bin.

We find no evidence for discontinuity in either the 2D or the 1D histograms.
Thus no bimodality is seen in the radio-loudness distribution. Our subsamples cover the same radio-loudness range at the regime where previous studies have reported bimodality using mixed sample of narrow- and broad-line galaxies ($\log {\cal R} \sim -6$ to $-2$, corresponding to ${\cal R}_K \sim 1$ to 10,000), indicating that this continuous distribution reflects the intrinsic nature of the AGN population accreting at lower to moderate accretion rates, in the local Universe rather than selection effects.
%\comment{Clearly, no bimodality is seen in the radio-loudness distribution. }

\subsection{Physical drivers of radio loudness in AGN}

%------------------------------- Figure -------------------------------%
\begin{figure*}[!ht]
  \centering
  % [trim={left bottom right top},clip]
  \includegraphics[width=\textwidth, trim=20 20 220 10]{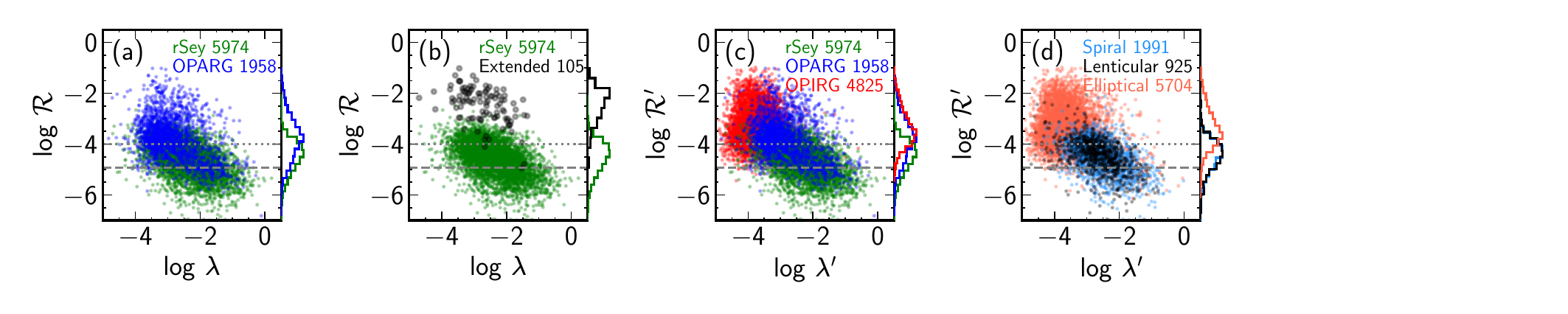}
  \caption{
  Scatter plots of radio loudness ($\mathcal{R}$) versus Eddington ratio ($\lambda$) for different subsamples: (a) radio Seyfert and OPARG; (b) radio Seyfert and extended sources; (c) radio Seyfert, OPARG, and OPIRG; and (d) by optical morphology (spiral, lenticular, elliptical).
  Extended sources are those classified with radio morphologies FRI, FRII, FRI/II, OI, OII, Z, X, DD, WAT, NAT, or HT in the ROGUE catalogues.
  Subsample sizes are indicated in each panel. 
  Marginal 1D histograms of radio loudness for each subsample are shown on the right hand side of each panel.
  Reference lines and primed and unprimed quantities are as in Fig.~\ref{fig:Rlambda2D}.
  %\comment{\sout{Values with and without a prime indicate \Lbol\ derived solely from \oiii\ or from a mix of \oiii\ (for OPARG and Seyfert) and $W3$ (for OPIRG) luminosities, respectively.}} 
  %\comment{\sout{The dashed line corresponds to the classical radio-loudness threshold of \citet{Kellermann.etal.1989a}, and the dotted line to the division from \citet{Gupta2018}; both are shown using our convention.}}
  }
  \label{fig:Rlambda}
\end{figure*}
% ------------------------------- Figure -------------------------------%

%------------------------------- Figure -------------------------------%
\begin{figure*}[!ht]
  \centering
  % [trim={left bottom right top},clip]
  \includegraphics[width=\textwidth, trim=20 20 220 10]{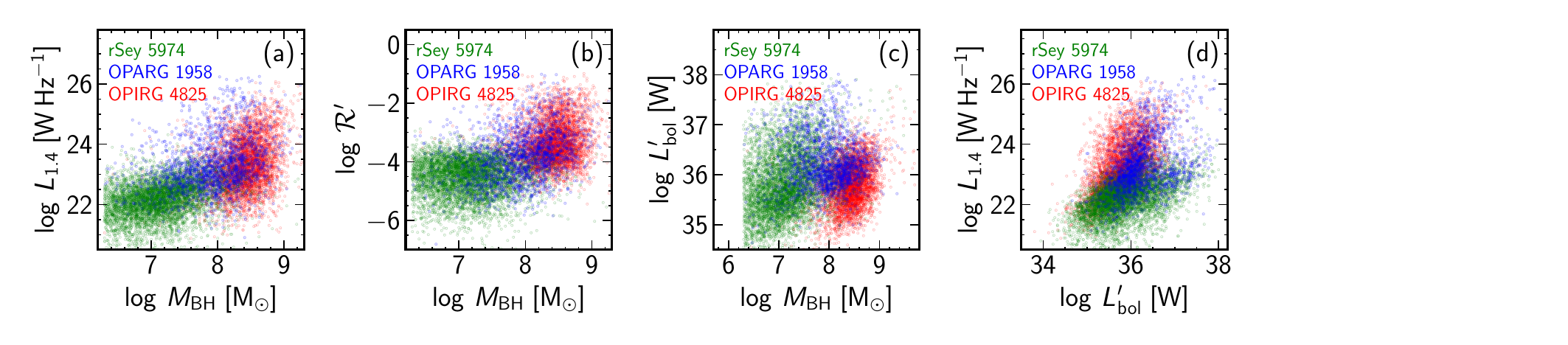}
  \caption{All panels show radio Seyferts, OPARGs, and OPIRGs (sample sizes in legends). (a) SF-corrected radio luminosity, (b) radio loudness, and (c) bolometric luminosity versus black hole mass. (d) Radio luminosity as a function of bolometric luminosity. (e) Radio size (measured from FIRST) as a function of black hole mass.}
  \label{fig:MBH}
\end{figure*}
% ------------------------------- Figure -------------------------------%

In Figure~\ref{fig:Rlambda}, we present the distribution of the radio-loudness parameter as a function of the Eddington ratio for different subsamples. On the righthand side of each panel we also show the projected 1D histogram for $\log {\cal R}$ for each subsample.
The majority of our sources accrete at moderate-to-low rate ($\log \lambda < -2$), within the range where radiatively inefficient accretion flows are expected to form. It should however be noted that radio Seyferts and OPARGs extend up to $\log \lambda \lesssim 0$, \ie into the regime that allows for the formation of standard geometrically thin, optically thick (Shakura–Sunyaev) accretion disks.

In panel (a) of Figure~\ref{fig:Rlambda}, we see that the radio Seyferts and OPARGs, \ie the two subsamples with accretion-related luminosities robustly derived from line emission, form a continuous distribution, characterized by a general anti-correlation in the $\lambda$--$\mathcal{R}$ plane, albeit with a significant spread. Specifically, for a given value of the Eddington ratio within the range $-4 < \log \lambda < -1$, the values of $\mathcal{R}$ span about two orders of magnitude. Moreover, the radio Seyfert and OPARG subsamples largely overlap, with the exception of a small number of OPARGs that reach the highest values of radio-loudness. 

Panel (b) of Figure~\ref{fig:Rlambda} shows that extended sources are concentrated in the high-$\mathcal{R}$ tail of the OPARGs. In other words, radio-compact galaxies (\ie unresolved in NVSS and FIRST) are indistinguishable from radio Seyferts in the $\lambda$--$\mathcal{R}$ plane, while only radio galaxies with extended radio structures reach much higher values of $\mathcal{R}$. Compact radio galaxies are generally excluded from population studies, as their identification is non-trivial. Most work therefore focuses on the `classical' extended FR I and FR II sources, which leads to an apparent bimodality in the radio-loudness distribution: extended radio galaxies appear more radio-loud than radio Seyferts at a given Eddington ratio (see \eg \citealt{Sikora.Stawarz.Lasota.2007a}). Our results show, however, that most radio galaxies are compact with moderate $\mathcal{R}$ values, with only a minority having extended structures and high radio loudness. When these compact galaxies are taken into account the apparent bimodality between radio Seyferts and radio galaxies disappears.
%When the full population is considered, rather than only the extended systems, the apparent bimodality between radio Seyferts and radio galaxies disappears.

Panel (c) of Figure~\ref{fig:Rlambda} shows that OPIRGs, \ie radio galaxies lacking prominent optical signatures of central activity, have the lowest accretion rates ($\log \lambda < -3$). 
The wider $\sim 3$~dex spread in $\mathcal{R}$ for OPIRGs may arise from deriving \Lbol from $W3$, as the AGN power can be overestimated due the contribution from the stellar populations of the galaxies to the $W3$ band.
Considering all three subsamples together, the Spearman correlation coefficient between $x = \log \cal{R}^\prime$ and $y = \log \lambda^\prime$ is $\rho_{xy} = -0.638$, $-0.499$ for radio Seyferts, $-0.331$ for OPARGs and $-0.128$ for OPIRGs. 
To account for the $\cal{R}$ and $\lambda$ shared dependence on \Lbol, we computed the partial correlation coefficient $\rho_{xy|z}$ controlling for $z = \log \Lbol^\prime$ \citep[see \eg eq.~1 in][]{Wild.etal.2025a}. For all subsamples taken together, $\rho_{xy|z} = -0.636$, $-0.234$ for radio Seyferts, $-0.293$ for OPARGs and $-0.196$ for OPIRGs. Thus the anticorrelation between $\cal{R}$ and $\lambda$ remains even after accounting for the effect of $\Lbol$.

Panel (d) of Figure~\ref{fig:Rlambda} shows that, when host morphology is considered, a stronger contrast between high and low-$\mathcal{R}$ sources emerges. Spiral and lenticular galaxies, predominantly hosting radio Seyferts and some OPARGs, lie in the lower-right region (lower $\mathcal{R}$, higher $\lambda$) and exhibit a clear $\lambda$–$\mathcal{R}$ anticorrelation. Ellipticals, hosting OPIRGs and most OPARGs, instead form a broad, featureless distribution at low $\lambda$ and mid-to-high $\mathcal{R}$.

Figure~\ref{fig:MBH} shows our subsamples in different \MBH versus luminosities or luminosity versus luminosity planes. In panel (a) we see that Seyfert galaxies are found in systems with relatively low black hole masses and low radio luminosities. 
OPIRGs are associated with the most massive black holes and represent the most radio-luminous objects in our sample. OPARGs occupy an intermediate regime, typically hosting intermediate black hole masses, while their radio luminosities can extend up to the ranges observed in OPIRGs.

In Figure~\ref{fig:MBH}(b) we show $\mathcal{R}$ as a function of \MBH. A saturation in radio loudness is clearly evident for Seyfert galaxies, with OPIRGs again occupying the upper corner of the plot, characterized by higher black hole masses and higher radio luminosities than Seyferts. However, no clear separation is observed: OPARGs populate the intermediate region between OPIRGs and Seyferts, showing some indication of a correlation between \MBH and radio loudness.

As shown in panel (c) of Figure~\ref{fig:MBH}, the bolometric accretion-related luminosities of our sources span a range of only about two orders of magnitude. In contrast, Figure~\ref{fig:MBH}(d) shows that for any given bolometric luminosity, the corresponding radio luminosity extends across nearly four to five orders of magnitude. The broad spread in the data results from three distinct clouds of points corresponding to radio Seyferts, OPARGs, and OPIRGs, with each cloud concentrated at progressively higher radio luminosities.
The comparatively narrow range in bolometric output and the wider dispersion in radio power highlight that the radio-loudness parameter is driven primarily by variations in radio luminosity rather than bolometric luminosity.

Here we note that the observed distributions presented in Figures~\ref{fig:Rlambda} and~\ref{fig:MBH} accurately reflect the actual distribution of sources within the considered redshift range ($z < 0.4$), given that they effectively captures the bulk of the studied populations, as noted in section \ref{sec:fluxlimit}. If the investigated sources undergo strong cosmological evolution, these distributions may appear substantially distinct, however such evolution is not expected as was reported in \cite{Smolcic2009,Padovani.etal.2015a} .

\subsection{Understanding the observed scaling relations}

The total power of a jet extracted via the Blandford–Znajek (BZ) mechanism \citep{Blandford.Znajek.1977a}, $P_j$, can be expressed as a function of the dimensionless magnetic flux parameter $0 < \varphi \leq 1$ (i.e., the magnetic flux threading the BH horizon, normalized to its maximum value corresponding to the `magnetically arrested disk' regime), the dimensionless BH spin $0 < a \leq 1$, and the mass accretion rate $\dot{M}_{\rm acc}$ \citep{Tchekhovskoy.Narayan.McKinney.2011a}:
\begin{equation}
P_j \simeq F\!(a) \, \varphi^2 \, \dot{M}_{\rm acc} c^2,
\end{equation}
where $F(a)$ is a function of the black hole spin. In Appendix~\ref{sec:appendixA}, we provide a simple derivation of this formula, along with the explicit form of the function $F(a)$.

Only a small fraction of this total jet power is radiated away at radio frequencies. In the literature, various scaling relations between $P_j$ and the observed radio luminosity $L_r$ have been discussed \citep{Willott1999, Cavagnolo2010, Sullivan2011, G&Sh2016}, typically consistent with a simple parametrization:
\begin{equation}
P_j \propto L_r^{\alpha} \quad \textrm{with} \quad 0.5 \leq \alpha \leq 1.0 \, ,
\end{equation}
which we adopt in the discussion below.

Moreover, the mass accretion rate can be expressed in terms of the Eddington parameter $\lambda$. From general considerations of accretion disk theory \citep{N98}, it follows that (see Appendix~\ref{sec:appendixB}):
\begin{equation}
    \lambda \simeq \left\{ 
  \begin{array}{ c l }
   \dot{m}^2/\dot{m}_{\rm cr}  & \quad \textrm{if } \quad \lambda < \dot{m}_{\rm cr}, \\
   \dot{m}  & \quad \textrm{if } \quad \lambda \geq \dot{m}_{\rm cr}, \\
  \end{array}
\right. 
\end{equation}
where $\dot{m}$ is the mass accretion rate normalized to Eddington units, $\dot{m} \equiv \dot{M}_{\rm acc} /\dot{M}_{\rm Edd}$ with $\dot{M}_{\rm Edd} \propto M_{\rm BH}$, and $\dot{m}_{\rm cr}$ is the critical dimensionless value separating the regime of radiatively inefficient accretion flows from standard Shakura–Sunyaev disks.

With the above considerations the expected scaling between the radio (jet related) and bolometric (accretion related) luminosity becomes:
\begin{equation}
    L_r\propto\begin{cases}
    \quad [\varphi^2 F\!(a)]^{1/\alpha}\,L_\mathrm{bol}^{1/2\alpha}\,M_\mathrm{BH}^{1/2\alpha}   & \quad \text{for }\quad \lambda<\dot{m}_{cr},
    \\
    \quad [\varphi^2F\!(a)]^{1/\alpha}\,L_\mathrm{bol}^{1/\alpha}       & \quad \text{for }\quad \lambda\geq\dot{m}_{cr}.
    \end{cases}
    \label{eq:radiolumalpha}
\end{equation}
The radio-loudness parameter, $\mathcal{R} = L_r/L_{\rm bol}$, on the other hand, becomes
\begin{equation}
\mathcal{R} \propto
    \begin{cases}
    \quad [\varphi^2 F\!(a)]^{1/\alpha}\,
    \lambda^{(0.5-\alpha)/\alpha}\,M_\mathrm{BH}^{(1-\alpha)/\alpha}     & \quad \text{for }\quad \lambda<\dot{m}_{cr},
   \\
    \quad [\varphi^2 F\!(a)]^{1/\alpha}\,
    \lambda^{(1-\alpha)/\alpha}\,M_\mathrm{BH}^{(1-\alpha)/\alpha}        & \quad \text{for }\quad \lambda\geq\dot{m}_{cr}.
    \end{cases}
    \label{eq:radioloudalpha}
\end{equation}

%\lukasz{\ojo [I did not look at the rest of the paper from this point (except of the Appendices which are now revised).]}

%\comment{\ojo. this does not look like an appendix to me. Could we manage to incorporate it in the main text?}

%As was noted in the text the difficulty in understanding the observed dependence of radio luminosity on both BH mass and disk bolometric luminosity is in the fact that this relation is given only for the jet power. The exact formula by which jet power varies with radio luminosity is not fully constrained and various scaling relations exist in the literature. %\copypaste
%------------------------------- Figure -------------------------------%
\begin{figure*}[!ht]
  \centering
  % [trim={left bottom right top},clip]
  \includegraphics[width=0.75\textwidth, trim=20 20 420 20]{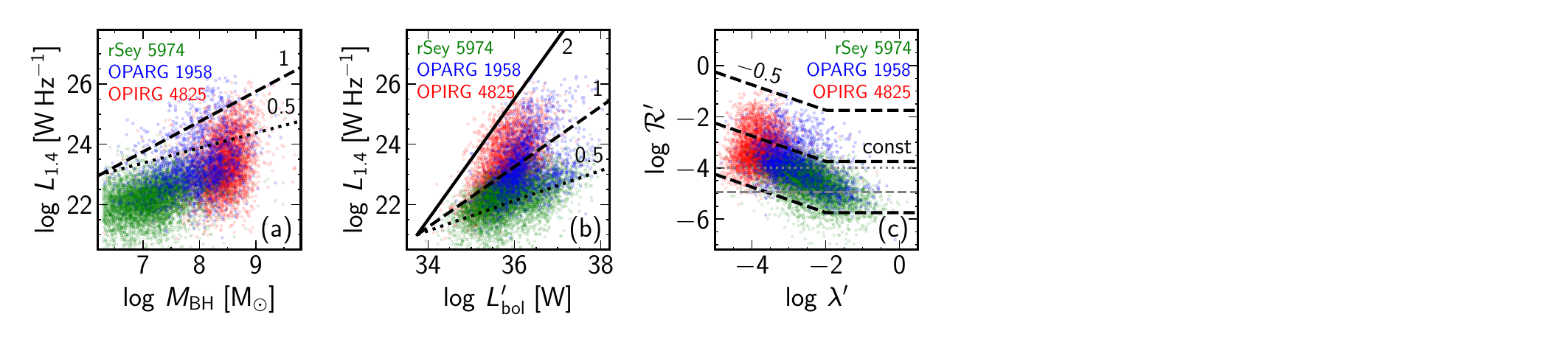}
  \caption{All panels show radio Seyferts, OPARGs, and OPIRGs (sample sizes in legends). Lines are labeled with their respective scaling relation exponents.
  (a) SF-corrected radio luminosity versus black hole mass (as in Fig.~\ref{fig:MBH}a). Overplotted lines show $\Lrad \propto \MBH$ (dashed) and $\Lrad \propto \MBH^{1/2}$ (dotted line); see eq.~\ref{eq:radiolumalpha}.
  (b) Radio luminosity as a function of bolometric luminosity (as in Fig.~\ref{fig:MBH}d). Overplotted lines show $\Lrad \propto \Lbol^2$ (filled), $\Lrad \propto \Lbol$ (dashed) and $\Lrad \propto \Lbol^{1/2}$ (dotted line); see eq.~\ref{eq:radiolumalpha}.
  (c) Radio loudness versus Eddington ratio (as in Fig.~\ref{fig:Rlambda}c). Overplotted dashed lines show, for three arbitrary normalizations, ${\cal{R}} \propto \lambda^{-1/2}$ up to a critical Eddington ratio of $\lambda = 10^{-2}$, above which ${\cal R}$ remains constant (see eq.~\ref{eq:radioloudalpha}).
  }
  \label{fig:append}
\end{figure*}
% ------------------------------- Figure -------------------------------%
In Fig.~\ref{fig:append} we show the radio luminosity as the function of BH mass (panel a) and bolometric luminosity (panel b) and radio-loudness versus Eddington ratio (panel c).
The functional form by which the radio-loudness depend on these parameters was given by Equations {\ref{eq:radiolumalpha} and \ref{eq:radioloudalpha}}.  In Figure~\ref{fig:append} we attempt to constrain $\alpha$ parameter by showing the lines corresponding to the scaling with $\alpha=0.5$ and $\alpha=1.0$. 
\iffalse \ania{I think maybe we don't need to keep this text}
In particular one can write explicitly the scaling corresponding to $\alpha=0.5$ and $\alpha=1.0$:
\begin{equation}
\label{eq:L-Lbol-MBH}
    \begin{cases}
    \quad \text{for }\quad \alpha=0.5: 
         \quad\ L_r\propto 
         \begin{cases}
             \quad L_\mathrm{bol}\,M_\mathrm{BH} & \quad \text{for } \quad \lambda <  \dot{m}_{cr}, \\
             \quad L_\mathrm{bol}^2              & \quad \text{for } \quad \lambda\geq\dot{m}_{cr}; \\
        \end{cases}\\
     \\
       % L_\mathrm{bol}^2         & \quad \text{for }\quad \dot{m}\geq\dot{m}_{cr} \\\\
    \quad \text{for }\quad \alpha=1.0: 
         \quad\ L_r\propto 
         \begin{cases}
             \quad L_\mathrm{bol}^{1/2}\,M_\mathrm{BH}^{1/2} & \quad \text{for }\quad \lambda <  \dot{m}_{cr}, \\
             \quad L_\mathrm{bol}                            & \quad \text{for }\quad \lambda\geq\dot{m}_{cr}. \\
         \end{cases}
    \end{cases}
\end{equation}
\ania{Rest we can keep}
\fi

The expected trends are indeed present in panels (a) and (b) of Figure~\ref{fig:append}, with the scatter corresponding to the variation in $[\varphi^2\,F(a)]^{\,1/\alpha}$.
However, a more robust interpretation is hampered by the fact that our subsamples, Seyferts, OPARGs and OPIRGs, differ in their distributions of $\lambda$ and \MBH. As a result, a simple comparison of the observed luminosity--luminosity trends with the expected $L_r$--$L_\mathrm{bol}$ scaling alone should be considered with caution, especially as a positive luminosity--luminosity correlation in a sample of cosmologically distant sources may not be intrinsic to the population but instead induced by flux limits.

In panel (c) of Figure~\ref{fig:append} we plot the radio-loudness versus Eddington ratio for our sample. An overall anti-correlation trend is apparent, which in fact is characteristic for the general AGN population (also present in \citealp{Sikora.Stawarz.Lasota.2007a}), and this favors a $\alpha>0.5$  scaling (unless \MBH is substantially anti-correlated with $\lambda$). 
Indeed, a closer look at this figure favors $\alpha\approx1$, which is in good agreement with the $\alpha\approx0.9$ obtained by \cite{Willott1999}, the dashed lines marking the relation expected for $\alpha=1$ and $\dot{m}_{\,cr}=10^{-2}$, with three arbitrary normalizations differing successively by a factor of 100. We note that distribution of radio-loudness does not explicitly depend on black hole mass, given that both model parameters, the dimensionless magnetic flux $\varphi$ and BH spin $a$ are independent on $M_\mathrm{BH}$.
This shows that, at fixed $\lambda$, the full observed spread in $\mathcal{R}$ can be explained solely by variations in $[\varphi^2 F(a)]$.

%This on turn implies that, 
The implication of those relations is that, while at higher Eddington ratio ($\lambda\geq\dot{m}_{cr}$) the radio luminosity is a linear function of the bolometric luminosity, the radio loudness parameter is expected to be flat, with the scatter resulting from the variation of the $[\varphi^2\,F(a)]$ term. At lower Eddington ratio ($\lambda<\dot{m}_{cr}$) this dependence becomes more complex, with radio luminosity varying as $L_r\propto\sqrt{L_\mathrm{bol}\,M_\mathrm{BH}}$ and radio loudness being a more simple function of Eddington ratio varying as ${\cal R}\propto 1/\sqrt{\lambda}$, with the scatter again resulting from the $[\varphi^2\,F(a)]$ term.\\

Let us now consider the expected scatter from this relation, and whether it can account for the variations observed in our sample.
Unlike the magnetic flux, the spin of a black hole can be constrained through astrophysical observations. However, this is possible only for a small number of objects, and the results are subject to observational biases and strong dependence on the method used. Current spin estimates for a limited sample tend to favor high spin values \citep{Soares2020, Reynolds2021}, consistent with a unimodal spin distribution.

To infer spin information for a much larger population of black holes, we must turn to cosmological simulations. In HORIZON-AGN \citep{Dubois2014}, initially non-spinning (Schwarzschild) black holes evolve through gas accretion and mergers, producing, by the present epoch, a continuous spin distribution spanning $a \sim 0.6$–$1.0$ \citep{Beckmann2024}. More recently, the NEWHORIZON simulation \citep{Dubois2021} investigated intermediate-mass and massive black holes, explicitly including spin-down from BZ feedback. Despite the simulation’s limited volume, this feedback reduces spins only slightly, leaving massive black holes mostly unaffected and preserving a continuous overall distribution \citep{Beckmann2025}.

Taking into account both observational constraints and numerical simulations, we conclude that spins in the range $a \sim 0.6$--$1.0$ are likely typical for the massive black holes considered in this work. However, variations within this interval can explain at most a factor of $\sim 7$ in the observed radio luminosities and radio loudness. The additional spread is more plausibly attributed to differences in the magnetic flux.

To estimate the required variation in the dimensionless magnetic flux ($\varphi$), we note that a change in $\varphi$ by a factor of $\sim 4$ across a given galaxy population, combined with independent scatter in black hole spins within the range $0.6 < a < 1.0$, would cause the term [$\varphi^2,F(a)$] to vary by about $\sim 2$ dex. 
This level of variation can account for the trends observed in the radio Seyfert sample as well as in the bulk of the OPARG/compact-radio subsamples. However, the OPARG/extended-radio and OPIRG sources are considerably more radio-loud, spanning over 4 dex in radio-loudness at a fixed $\lambda$. Explaining this spread within our scaling relations would therefore require a much larger variation in $\varphi$ across these populations.

We further note that extended radio sources in the OPARG and OPIRG categories are preferentially located on the upper branch of the radio-loudness versus Eddington ratio diagram (see Figure~\ref{fig:Rlambda}). This suggests that long jet duty cycles characteristic of these systems are necessary for galaxies to reach the most extreme radio-loudness parameter values.

\section{Conclusions}

We analyzed the radio-loudness distribution of a large, homogeneous sample of radio galaxies from the local Universe ($z<0.4$) selected from the SDSS MGS and present in the ROGUE I and II catalogues. 
%in  galaxies from the the ROGUE I and II catalogues of radio sources (\citealp{KozielWierzbowska.Goyal.Zywucka.2020a}; 2025, \comment{to be submitted}) that belong to the Main Galaxy Sample  of the Sloan Digital Sky Survey (SDSS; 
%In this work we analyzed the radio-loudness distribution of a large, homogeneously selected sample of radio galaxies from the ROGUE I and II catalogues. 
%\comment{Galaxies have been selected from the SDSS Main Galaxy Sample cross-matched with FIRST and NVSS.} 
All sources, regardless of their radio extent or luminosity, were included. By considering both compact and extended systems and accounting for the contribution of star formation to the radio emission, we inferred the relation between radio-loudness, accretion rate, and host galaxy properties. Our subsamples of radio Seyferts, OPIRGs and OPARGs were primarily defined by their positions in the $D_n(4000)$ versus $L_{1.4}/M_\star$ plane, their \Ha equivalent widths, and their positions in the BPT diagram.
Our conclusions are as follows.

\begin{itemize}

\item Since  our sample includes compact radio sources, which constitute the majority of radio galaxies, the distribution of radio-loudness appears smooth and definitely not bi-modal, as found in some previous studies.

%we do not }Our study does not find compelling evidence for a bimodal distribution of the radio-loudness parameter in the whole population of investigated radio sources.

\item The highest $\mathcal{R}$ values are reached almost exclusively by radio galaxies with extended radio structures, such as FRI, FRII, FRI/II and others.

\item The radio-loudness parameter $\mathcal{R}$ is anti-correlated with the Eddington ratio $\lambda$. 

\item We find a large spread in the radio-loudness parameter $\mathcal{R}$ values of two to three orders of magnitude at any given accretion rate. The $\mathcal{R}$ scatter increases with decreasing $\lambda$, especially at the high-$\mathcal{R}$ tail of the distribution for both OPIRGs and OPARGs.

\item Spirals and lenticulars occupy the high-$\lambda$, low-$\mathcal{R}$ region, while ellipticals lie at lower $\lambda$ and higher $\mathcal{R}$ and exhibit a larger scatter.

\item Radio Seyferts, OPARGs and OPIRGs form a sequence of increasing \MBH with a large overlap among subsamples. Radio Seyferts show no correlation between radio loudness with \MBH, %\comment{\ojo Ania can you explain why? }
while OPARGs and OPIRGs show a weak trend between \MBH and $\mathcal{R}$.

\item By applying theoretical models, we derived scaling relations that relate radio luminosity and radio loudness to bolometric (accretion-related) luminosity and black hole mass.

\item Within the framework of the derived scaling relations, the observed spread of up to two orders of magnitude in both radio luminosity and radio loudness can be attributed to variations in the dimensionless magnetic flux $\varphi$ by only a factor of $\sim 4$, under the assumption that black hole spins are in the range $a = 0.6$--$1.0$.

\end{itemize}

\begin{acknowledgements}

This work was done within the framework of the research project no.\ 2021/43/B/ST9/03246 financed by the National Science Centre/Narodowe Centrum Nauki. 
AW was supported by the GACR grant 21-13491X.
NVA acknowledges support of Conselho Nacional de Desenvolvimento Cient\'{i}fico e Tecnol\'{o}gico (CNPq). 
Funding for the creation and distribution of the SDSS Archive has been provided by the Alfred P. Sloan Foundation, the Participating Institutions, the National Aeronautics and Space Administration, the National Science Foundation, the U.S. Department of Energy, the Japanese Monbukagakusho, and the Max Planck Society. The SDSS Web site is \url{http://www.sdss.org/}.
The Participating Institutions are The University of Chicago, Fermilab, the Institute for Advanced Study, the Japan Participation Group, The Johns Hopkins University, the Max-Planck-Institute for Astronomy (MPIA), the Max-Planck-Institute for Astrophysics (MPA), New Mexico State University, Princeton University, the United States Naval Observatory, and the University of Washington.

This research made use of Astropy,\footnote{Astropy Python package: \url{http://www.astropy.org}} a community-developed core Python package for Astronomy \citep{AstropyCollaboration.etal.2013a, AstropyCollaboration.etal.2018a}.

\end{acknowledgements}

\bibliographystyle{aa}
\bibliography{references}

@article{Wild.etal.2025a,
	author = {{Wild}, Vivienne and {Asari}, Natalia Vale and {Rowlands}, Kate and {Ellison}, Sara L. and {Leung}, Ho-Hin and {Tremonti}, Christy},
	doi = {10.33232/001c.128125},
	journal = {The Open Journal of Astrophysics},
	month = jan,
	pages = {3},
	title = {{The infrared luminosity of retired and post-starburst galaxies: A cautionary tale for star formation rate measurements}},
	volume = {8},
	year = 2025}

@ARTICLE{Koziel2017,
       author = {{Kozie{\l}-Wierzbowska}, D. and {Vale Asari}, N. and {Stasi{\'n}ska}, G. and {Sikora}, M. and {Goettems}, E.~I. and {W{\'o}jtowicz}, A.},
        title = "{What Distinguishes the Host Galaxies of Radio-loud and Radio-quiet AGNs?}",
      journal = {\apj},
         year = 2017,
        month = sep,
       volume = {846},
       number = {1},
          eid = {42},
        pages = {42},
          doi = {10.3847/1538-4357/aa8326},
archivePrefix = {arXiv},
       eprint = {1709.09912},
 primaryClass = {astro-ph.GA},
       adsurl = {https://ui.adsabs.harvard.edu/abs/2017ApJ...846...42K}
}

@ARTICLE{Zheng2020,
       author = {{Zheng}, X.~C. and {R{\"o}ttgering}, H.~J.~A. and {Best}, P.~N. and {van der Wel}, A. and {Hardcastle}, M.~J. and {Williams}, W.~L. and {Bonato}, M. and {Prandoni}, I. and {Smith}, D.~J.~B. and {Leslie}, S.~K.},
        title = "{Link between radio-loud AGNs and host-galaxy shape}",
      journal = {\aap},
         year = 2020,
        month = dec,
       volume = {644},
          eid = {A12},
        pages = {A12},
          doi = {10.1051/0004-6361/202038646},
archivePrefix = {arXiv},
       eprint = {2010.07851},
 primaryClass = {astro-ph.GA},
       adsurl = {https://ui.adsabs.harvard.edu/abs/2020A&A...644A..12Z}
}

@ARTICLE{Rankine2021,
       author = {{Rankine}, Amy L. and {Matthews}, James H. and {Hewett}, Paul C. and {Banerji}, Manda and {Morabito}, Leah K. and {Richards}, Gordon T.},
        title = "{Placing LOFAR-detected quasars in C IV emission space: implications for winds, jets and star formation}",
      journal = {\mnras},
         year = 2021,
        month = apr,
       volume = {502},
       number = {3},
        pages = {4154-4169},
          doi = {10.1093/mnras/stab302},
archivePrefix = {arXiv},
       eprint = {2101.12635},
 primaryClass = {astro-ph.GA},
       adsurl = {https://ui.adsabs.harvard.edu/abs/2021MNRAS.502.4154R}
}

@ARTICLE{Macfarlane2021,
       author = {{Macfarlane}, C. and {Best}, P.~N. and {Sabater}, J. and {G{\"u}rkan}, G. and {Jarvis}, M.~J. and {R{\"o}ttgering}, H.~J.~A. and {Baldi}, R.~D. and {Calistro Rivera}, G. and {Duncan}, K.~J. and {Morabito}, L.~K. and {Prandoni}, I. and {Retana-Montenegro}, E.},
        title = "{The radio loudness of SDSS quasars from the LOFAR Two-metre Sky Survey: ubiquitous jet activity and constraints on star formation}",
      journal = {\mnras},
         year = 2021,
        month = oct,
       volume = {506},
       number = {4},
        pages = {5888-5907},
          doi = {10.1093/mnras/stab1998},
archivePrefix = {arXiv},
       eprint = {2107.09141},
 primaryClass = {astro-ph.GA},
       adsurl = {https://ui.adsabs.harvard.edu/abs/2021MNRAS.506.5888M}
}

@ARTICLE{Hardcastle2020,
       author = {{Hardcastle}, M.~J. and {Croston}, J.~H.},
        title = "{Radio galaxies and feedback from AGN jets}",
      journal = {\nar},
         year = 2020,
        month = jun,
       volume = {88},
          eid = {101539},
        pages = {101539},
          doi = {10.1016/j.newar.2020.101539},
archivePrefix = {arXiv},
       eprint = {2003.06137},
 primaryClass = {astro-ph.HE},
       adsurl = {https://ui.adsabs.harvard.edu/abs/2020NewAR..8801539H}
}

@ARTICLE{Saika2022,
       author = {{Saikia}, D.~J.},
        title = "{Jets in radio galaxies and quasars: an observational perspective}",
      journal = {Journal of Astrophysics and Astronomy},
         year = 2022,
        month = dec,
       volume = {43},
       number = {2},
          eid = {97},
        pages = {97},
          doi = {10.1007/s12036-022-09863-2},
archivePrefix = {arXiv},
       eprint = {2206.05803},
 primaryClass = {astro-ph.GA},
       adsurl = {https://ui.adsabs.harvard.edu/abs/2022JApA...43...97S}
}

@ARTICLE{Gupta2018,
       author = {{Gupta}, Maitrayee and {Sikora}, Marek and {Rusinek}, Katarzyna and {Madejski}, Greg M.},
        title = "{Comparison of hard X-ray spectra of luminous radio galaxies and their radio-quiet counterparts}",
      journal = {\mnras},
         year = 2018,
        month = nov,
       volume = {480},
       number = {3},
        pages = {2861-2871},
          doi = {10.1093/mnras/sty2043},
archivePrefix = {arXiv},
       eprint = {1808.07170},
 primaryClass = {astro-ph.HE},
       adsurl = {https://ui.adsabs.harvard.edu/abs/2018MNRAS.480.2861G}
}

@ARTICLE{Smolcic2009,
       author = {{Smol{\v{c}}i{\'c}}, V. and {Zamorani}, G. and {Schinnerer}, E. and {Bardelli}, S. and {Bondi}, M. and {B{\^\i}rzan}, L. and {Carilli}, C.~L. and {Ciliegi}, P. and {Elvis}, M. and {Impey}, C.~D. and {Koekemoer}, A.~M. and {Merloni}, A. and {Paglione}, T. and {Salvato}, M. and {Scodeggio}, M. and {Scoville}, N. and {Trump}, J.~R.},
        title = "{Cosmic Evolution of Radio Selected Active Galactic Nuclei in the Cosmos Field}",
      journal = {\apj},
         year = 2009,
        month = may,
       volume = {696},
       number = {1},
        pages = {24-39},
          doi = {10.1088/0004-637X/696/1/24},
archivePrefix = {arXiv},
       eprint = {0901.3372},
 primaryClass = {astro-ph.CO},
       adsurl = {https://ui.adsabs.harvard.edu/abs/2009ApJ...696...24S}
}

@ARTICLE{Singh2018,
       author = {{Singh}, Veeresh and {Chand}, Hum},
        title = "{Investigating kpc-scale radio emission properties of narrow-line Seyfert 1 galaxies}",
      journal = {\mnras},
         year = 2018,
        month = oct,
       volume = {480},
       number = {2},
        pages = {1796-1818},
          doi = {10.1093/mnras/sty1818},
archivePrefix = {arXiv},
       eprint = {1807.01945},
 primaryClass = {astro-ph.GA},
       adsurl = {https://ui.adsabs.harvard.edu/abs/2018MNRAS.480.1796S}
}

@ARTICLE{Xu99,
       author = {{Xu}, Chun and {Livio}, Mario and {Baum}, Stefi},
        title = "{Radio-loud and Radio-quiet Active Galactic Nuclei}",
      journal = {\aj},
         year = 1999,
        month = sep,
       volume = {118},
       number = {3},
        pages = {1169-1176},
          doi = {10.1086/301007},
archivePrefix = {arXiv},
       eprint = {astro-ph/9905322},
 primaryClass = {astro-ph},
       adsurl = {https://ui.adsabs.harvard.edu/abs/1999AJ....118.1169X}
}

@ARTICLE{Mahony12,
       author = {{Mahony}, Elizabeth K. and {Sadler}, Elaine M. and {Croom}, Scott M. and {Ekers}, Ronald D. and {Feain}, Ilana J. and {Murphy}, Tara},
        title = "{Is the Observed High-frequency Radio Luminosity Distribution of QSOs Bimodal?}",
      journal = {\apj},
         year = 2012,
        month = jul,
       volume = {754},
       number = {1},
          eid = {12},
        pages = {12},
          doi = {10.1088/0004-637X/754/1/12},
archivePrefix = {arXiv},
       eprint = {1205.2233},
 primaryClass = {astro-ph.CO},
       adsurl = {https://ui.adsabs.harvard.edu/abs/2012ApJ...754...12M}
}

@ARTICLE{Ivezic02,
       author = {{Ivezi{\'c}}, {\v{Z}}eljko and {Menou}, Kristen and {Knapp}, Gillian R. and {Strauss}, Michael A. and {Lupton}, Robert H. and {Vanden Berk}, Daniel E. and {Richards}, Gordon T. and {Tremonti}, Christy and {Weinstein}, Michael A. and {Anderson}, Scott and {Bahcall}, Neta A. and {Becker}, Robert H. and {Bernardi}, Mariangela and {Blanton}, Michael and {Eisenstein}, Daniel and {Fan}, Xiaohui and {Finkbeiner}, Douglas and {Finlator}, Kristian and {Frieman}, Joshua and {Gunn}, James E. and {Hall}, Pat B. and {Kim}, Rita S.~J. and {Kinkhabwala}, Ali and {Narayanan}, Vijay K. and {Rockosi}, Constance M. and {Schlegel}, David and {Schneider}, Donald P. and {Strateva}, Iskra and {SubbaRao}, Mark and {Thakar}, Aniruddha R. and {Voges}, Wolfgang and {White}, Richard L. and {Yanny}, Brian and {Brinkmann}, Jonathan and {Doi}, Mamoru and {Fukugita}, Masataka and {Hennessy}, Gregory S. and {Munn}, Jeffrey A. and {Nichol}, Robert C. and {York}, Donald G.},
        title = "{Optical and Radio Properties of Extragalactic Sources Observed by the FIRST Survey and the Sloan Digital Sky Survey}",
      journal = {\aj},
         year = 2002,
        month = nov,
       volume = {124},
       number = {5},
        pages = {2364-2400},
          doi = {10.1086/344069},
archivePrefix = {arXiv},
       eprint = {astro-ph/0202408},
 primaryClass = {astro-ph},
       adsurl = {https://ui.adsabs.harvard.edu/abs/2002AJ....124.2364I}
}

@ARTICLE{Kellermann16,
       author = {{Kellermann}, K.~I. and {Condon}, J.~J. and {Kimball}, A.~E. and {Perley}, R.~A. and {Ivezi{\'c}}, {\v{Z}}eljko},
        title = "{Radio-loud and Radio-quiet QSOs}",
      journal = {\apj},
         year = 2016,
        month = nov,
       volume = {831},
       number = {2},
          eid = {168},
        pages = {168},
          doi = {10.3847/0004-637X/831/2/168},
archivePrefix = {arXiv},
       eprint = {1608.04586},
 primaryClass = {astro-ph.GA},
       adsurl = {https://ui.adsabs.harvard.edu/abs/2016ApJ...831..168K}
}

@ARTICLE{Ho01,
       author = {{Ho}, Luis C. and {Peng}, Chien Y.},
        title = "{Nuclear Luminosities and Radio Loudness of Seyfert Nuclei}",
      journal = {\apj},
         year = 2001,
        month = jul,
       volume = {555},
       number = {2},
        pages = {650-662},
          doi = {10.1086/321524},
archivePrefix = {arXiv},
       eprint = {astro-ph/0102502},
 primaryClass = {astro-ph},
       adsurl = {https://ui.adsabs.harvard.edu/abs/2001ApJ...555..650H}
}

@ARTICLE{Tadhunter16,
       author = {{Tadhunter}, Clive},
        title = "{Radio AGN in the local universe: unification, triggering and evolution}",
      journal = {\aapr},
         year = 2016,
        month = jun,
       volume = {24},
       number = {1},
          eid = {10},
        pages = {10},
          doi = {10.1007/s00159-016-0094-x},
archivePrefix = {arXiv},
       eprint = {1605.08773},
 primaryClass = {astro-ph.GA},
       adsurl = {https://ui.adsabs.harvard.edu/abs/2016A&ARv..24...10T}
}

@article{Wright.etal.2010a,
	author = {{Wright}, E.~L. and {Eisenhardt}, P.~R.~M. and {Mainzer}, A.~K. and {Ressler}, M.~E. and {Cutri}, R.~M. and {Jarrett}, T. and {Kirkpatrick}, J.~D. and {Padgett}, D. and {McMillan}, R.~S. and {Skrutskie}, M. and {Stanford}, S.~A. and {Cohen}, M. and {Walker}, R.~G. and {Mather}, J.~C. and {Leisawitz}, D. and {Gautier}, III, T.~N. and {McLean}, I. and {Benford}, D. and {Lonsdale}, C.~J. and {Blain}, A. and {Mendez}, B. and {Irace}, W.~R. and {Duval}, V. and {Liu}, F. and {Royer}, D. and {Heinrichsen}, I. and {Howard}, J. and {Shannon}, M. and {Kendall}, M. and {Walsh}, A.~L. and {Larsen}, M. and {Cardon}, J.~G. and {Schick}, S. and {Schwalm}, M. and {Abid}, M. and {Fabinsky}, B. and {Naes}, L. and {Tsai}, C.-W.},
	doi = {10.1088/0004-6256/140/6/1868},
	journal = {\aj},
	month = dec,
	pages = {1868-1881},
	title = {{The Wide-field Infrared Survey Explorer (WISE): Mission Description and Initial On-orbit Performance}},
	volume = 140,
	year = 2010}

@article{Asari.etal.2007a,
	author = {{Asari}, N.~V. and {Cid Fernandes}, R. and {Stasi{\'n}ska}, G. and {Torres-Papaqui}, J.~P. and {Mateus}, A. and {Sodr{\'e}}, L. and {Schoenell}, W. and {Gomes}, J.~M.},
	journal = {\mnras},
	month = oct,
	pages = {263-279},
	title = {{The history of star-forming galaxies in the Sloan Digital Sky Survey}},
	volume = 381,
	year = 2007}

@ARTICLE{Beckmann2025,
       author = {{Beckmann}, R.~S. and {Dubois}, Y. and {Volonteri}, M. and {Dong-Paez}, C.~A. and {Peirani}, S. and {Piotrowska}, J.~M. and {Martin}, G. and {Kraljic}, K. and {Devriendt}, J. and {Pichon}, C. and {Yi}, S.~K.},
        title = "{Black hole spin evolution across cosmic time from the NEWHORIZON simulation}",
      journal = {\mnras},
         year = 2025,
        month = jan,
       volume = {536},
       number = {2},
        pages = {1838-1856},
          doi = {10.1093/mnras/stae2595},
archivePrefix = {arXiv},
       eprint = {2410.02875},
 primaryClass = {astro-ph.HE},
       adsurl = {https://ui.adsabs.harvard.edu/abs/2025MNRAS.536.1838B}
}

@ARTICLE{Dubois2014,
       author = {{Dubois}, Yohan and {Volonteri}, Marta and {Silk}, Joseph},
        title = "{Black hole evolution - III. Statistical properties of mass growth and spin evolution using large-scale hydrodynamical cosmological simulations}",
      journal = {\mnras},
         year = 2014,
        month = may,
       volume = {440},
       number = {2},
        pages = {1590-1606},
          doi = {10.1093/mnras/stu373},
archivePrefix = {arXiv},
       eprint = {1304.4583},
 primaryClass = {astro-ph.CO},
       adsurl = {https://ui.adsabs.harvard.edu/abs/2014MNRAS.440.1590D}
}

@ARTICLE{Reynolds2021,
       author = {{Reynolds}, Christopher S.},
        title = "{Observational Constraints on Black Hole Spin}",
      journal = {\araa},
         year = 2021,
        month = sep,
       volume = {59},
        pages = {117-154},
          doi = {10.1146/annurev-astro-112420-035022},
archivePrefix = {arXiv},
       eprint = {2011.08948},
 primaryClass = {astro-ph.HE},
       adsurl = {https://ui.adsabs.harvard.edu/abs/2021ARA&A..59..117R}
}

@ARTICLE{Soares2020,
       author = {{Soares}, Gustavo and {Nemmen}, Rodrigo},
        title = "{Jet efficiencies and black hole spins in jetted quasars}",
      journal = {\mnras},
         year = 2020,
        month = jun,
       volume = {495},
       number = {1},
        pages = {981-991},
          doi = {10.1093/mnras/staa1241},
archivePrefix = {arXiv},
       eprint = {2005.00381},
 primaryClass = {astro-ph.HE},
       adsurl = {https://ui.adsabs.harvard.edu/abs/2020MNRAS.495..981S}
}

@ARTICLE{Dubois2021,
       author = {{Dubois}, Yohan and {Beckmann}, Ricarda and {Bournaud}, Fr{\'e}d{\'e}ric and {Choi}, Hoseung and {Devriendt}, Julien and {Jackson}, Ryan and {Kaviraj}, Sugata and {Kimm}, Taysun and {Kraljic}, Katarina and {Laigle}, Clotilde and {Martin}, Garreth and {Park}, Min-Jung and {Peirani}, S{\'e}bastien and {Pichon}, Christophe and {Volonteri}, Marta and {Yi}, Sukyoung K.},
        title = "{Introducing the NEWHORIZON simulation: Galaxy properties with resolved internal dynamics across cosmic time}",
      journal = {\aap},
         year = 2021,
        month = jul,
       volume = {651},
          eid = {A109},
        pages = {A109},
          doi = {10.1051/0004-6361/202039429},
archivePrefix = {arXiv},
       eprint = {2009.10578},
 primaryClass = {astro-ph.GA},
       adsurl = {https://ui.adsabs.harvard.edu/abs/2021A&A...651A.109D}
}

@article{AstropyCollaboration.etal.2013a,
	author = {{Astropy Collaboration} and {Robitaille}, T.~P. and {Tollerud}, E.~J. and {Greenfield}, P. and {Droettboom}, M. and {Bray}, E. and {Aldcroft}, T. and {Davis}, M. and {Ginsburg}, A. and {Price-Whelan}, A.~M. and {Kerzendorf}, W.~E. and {Conley}, A. and {Crighton}, N. and {Barbary}, K. and {Muna}, D. and {Ferguson}, H. and {Grollier}, F. and {Parikh}, M.~M. and {Nair}, P.~H. and {Unther}, H.~M. and {Deil}, C. and {Woillez}, J. and {Conseil}, S. and {Kramer}, R. and {Turner}, J.~E.~H. and {Singer}, L. and {Fox}, R. and {Weaver}, B.~A. and {Zabalza}, V. and {Edwards}, Z.~I. and {Azalee Bostroem}, K. and {Burke}, D.~J. and {Casey}, A.~R. and {Crawford}, S.~M. and {Dencheva}, N. and {Ely}, J. and {Jenness}, T. and {Labrie}, K. and {Lim}, P.~L. and {Pierfederici}, F. and {Pontzen}, A. and {Ptak}, A. and {Refsdal}, B. and {Servillat}, M. and {Streicher}, O.},
	journal = {\aap},
	month = oct,
	pages = {A33},
	title = {{Astropy: A community Python package for astronomy}},
	volume = 558,
	year = 2013}

@article{AstropyCollaboration.etal.2018a,
	author = {{Astropy Collaboration} and {Price-Whelan}, A.~M. and {Sip{\H o}cz}, B.~M. and {G{\"u}nther}, H.~M. and {Lim}, P.~L. and {Crawford}, S.~M. and {Conseil}, S. and {Shupe}, D.~L. and {Craig}, M.~W. and {Dencheva}, N. and {Ginsburg}, A. and {VanderPlas}, J.~T. and {Bradley}, L.~D. and {P{\'e}rez-Su{\'a}rez}, D. and {de Val-Borro}, M. and {Aldcroft}, T.~L. and {Cruz}, K.~L. and {Robitaille}, T.~P. and {Tollerud}, E.~J. and {Ardelean}, C. and {Babej}, T. and {Bach}, Y.~P. and {Bachetti}, M. and {Bakanov}, A.~V. and {Bamford}, S.~P. and {Barentsen}, G. and {Barmby}, P. and {Baumbach}, A. and {Berry}, K.~L. and {Biscani}, F. and {Boquien}, M. and {Bostroem}, K.~A. and {Bouma}, L.~G. and {Brammer}, G.~B. and {Bray}, E.~M. and {Breytenbach}, H. and {Buddelmeijer}, H. and {Burke}, D.~J. and {Calderone}, G. and {Cano Rodr{\'{\i}}guez}, J.~L. and {Cara}, M. and {Cardoso}, J.~V.~M. and {Cheedella}, S. and {Copin}, Y. and {Corrales}, L. and {Crichton}, D. and {D'Avella}, D. and {Deil}, C. and {Depagne}, {\'E}. and {Dietrich}, J.~P. and {Donath}, A. and {Droettboom}, M. and {Earl}, N. and {Erben}, T. and {Fabbro}, S. and {Ferreira}, L.~A. and {Finethy}, T. and {Fox}, R.~T. and {Garrison}, L.~H. and {Gibbons}, S.~L.~J. and {Goldstein}, D.~A. and {Gommers}, R. and {Greco}, J.~P. and {Greenfield}, P. and {Groener}, A.~M. and {Grollier}, F. and {Hagen}, A. and {Hirst}, P. and {Homeier}, D. and {Horton}, A.~J. and {Hosseinzadeh}, G. and {Hu}, L. and {Hunkeler}, J.~S. and {Ivezi{\'c}}, {\v Z}. and {Jain}, A. and {Jenness}, T. and {Kanarek}, G. and {Kendrew}, S. and {Kern}, N.~S. and {Kerzendorf}, W.~E. and {Khvalko}, A. and {King}, J. and {Kirkby}, D. and {Kulkarni}, A.~M. and {Kumar}, A. and {Lee}, A. and {Lenz}, D. and {Littlefair}, S.~P. and {Ma}, Z. and {Macleod}, D.~M. and {Mastropietro}, M. and {McCully}, C. and {Montagnac}, S. and {Morris}, B.~M. and {Mueller}, M. and {Mumford}, S.~J. and {Muna}, D. and {Murphy}, N.~A. and {Nelson}, S. and {Nguyen}, G.~H. and {Ninan}, J.~P. and {N{\"o}the}, M. and {Ogaz}, S. and {Oh}, S. and {Parejko}, J.~K. and {Parley}, N. and {Pascual}, S. and {Patil}, R. and {Patil}, A.~A. and {Plunkett}, A.~L. and {Prochaska}, J.~X. and {Rastogi}, T. and {Reddy Janga}, V. and {Sabater}, J. and {Sakurikar}, P. and {Seifert}, M. and {Sherbert}, L.~E. and {Sherwood-Taylor}, H. and {Shih}, A.~Y. and {Sick}, J. and {Silbiger}, M.~T. and {Singanamalla}, S. and {Singer}, L.~P. and {Sladen}, P.~H. and {Sooley}, K.~A. and {Sornarajah}, S. and {Streicher}, O. and {Teuben}, P. and {Thomas}, S.~W. and {Tremblay}, G.~R. and {Turner}, J.~E.~H. and {Terr{\'o}n}, V. and {van Kerkwijk}, M.~H. and {de la Vega}, A. and {Watkins}, L.~L. and {Weaver}, B.~A. and {Whitmore}, J.~B. and {Woillez}, J. and {Zabalza}, V. and {Astropy Contributors}},
	journal = {\aj},
	month = sep,
	pages = {123},
	title = {{The Astropy Project: Building an Open-science Project and Status of the v2.0 Core Package}},
	volume = 156,
	year = 2018}

@ARTICLE{Lumfuncellipticals,
       author = {{Im}, Myungshin and {Griffiths}, Richard E. and {Ratnatunga}, Kavan U. and {Sarajedini}, Vicki L.},
        title = "{Luminosity Functions of Elliptical Galaxies at Z <1.2}",
      journal = {\apjl},
         year = 1996,
        month = apr,
       volume = {461},
        pages = {L79},
          doi = {10.1086/310018},
archivePrefix = {arXiv},
       eprint = {astro-ph/9602041},
 primaryClass = {astro-ph},
       adsurl = {https://ui.adsabs.harvard.edu/abs/1996ApJ...461L..79I}
}

@ARTICLE{Ermash2013,
       author = {{Ermash}, A.~A.},
        title = "{The luminosity function of narrow-line Seyfert 1 galaxies based on SDSS data (DR7)}",
      journal = {Astronomy Reports},
         year = 2013,
        month = may,
       volume = {57},
       number = {5},
        pages = {317-326},
          doi = {10.1134/S106377291305003X},
archivePrefix = {arXiv},
       eprint = {1302.2955},
 primaryClass = {astro-ph.GA},
       adsurl = {https://ui.adsabs.harvard.edu/abs/2013ARep...57..317E}
}

@ARTICLE{Willott1999,
       author = {{Willott}, Chris J. and {Rawlings}, Steve and {Blundell}, Katherine M. and {Lacy}, Mark},
        title = "{The emission line-radio correlation for radio sources using the 7C Redshift Survey}",
      journal = {\mnras},
         year = 1999,
        month = nov,
       volume = {309},
       number = {4},
        pages = {1017-1033},
          doi = {10.1046/j.1365-8711.1999.02907.x},
archivePrefix = {arXiv},
       eprint = {astro-ph/9905388},
 primaryClass = {astro-ph},
       adsurl = {https://ui.adsabs.harvard.edu/abs/1999MNRAS.309.1017W}
}

@ARTICLE{Cavagnolo2010,
       author = {{Cavagnolo}, K.~W. and {McNamara}, B.~R. and {Nulsen}, P.~E.~J. and {Carilli}, C.~L. and {Jones}, C. and {B{\^\i}rzan}, L.},
        title = "{A Relationship Between AGN Jet Power and Radio Power}",
      journal = {\apj},
         year = 2010,
        month = sep,
       volume = {720},
       number = {2},
        pages = {1066-1072},
          doi = {10.1088/0004-637X/720/2/1066},
archivePrefix = {arXiv},
       eprint = {1006.5699},
 primaryClass = {astro-ph.CO},
       adsurl = {https://ui.adsabs.harvard.edu/abs/2010ApJ...720.1066C}
}

@ARTICLE{Sullivan2011,
       author = {{O'Sullivan}, E. and {Giacintucci}, S. and {David}, L.~P. and {Gitti}, M. and {Vrtilek}, J.~M. and {Raychaudhury}, S. and {Ponman}, T.~J.},
        title = "{Heating the Hot Atmospheres of Galaxy Groups and Clusters with Cavities: The Relationship between Jet Power and Low-frequency Radio Emission}",
      journal = {\apj},
         year = 2011,
        month = jul,
       volume = {735},
       number = {1},
          eid = {11},
        pages = {11},
          doi = {10.1088/0004-637X/735/1/11},
archivePrefix = {arXiv},
       eprint = {1104.2411},
 primaryClass = {astro-ph.CO},
       adsurl = {https://ui.adsabs.harvard.edu/abs/2011ApJ...735...11O}
}

@ARTICLE{G&Sh2016,
       author = {{Godfrey}, L.~E.~H. and {Shabala}, S.~S.},
        title = "{Mutual distance dependence drives the observed jet-power-radio-luminosity scaling relations in radio galaxies}",
      journal = {\mnras},
         year = 2016,
        month = feb,
       volume = {456},
       number = {2},
        pages = {1172-1184},
          doi = {10.1093/mnras/stv2712},
archivePrefix = {arXiv},
       eprint = {1511.06007},
 primaryClass = {astro-ph.GA},
       adsurl = {https://ui.adsabs.harvard.edu/abs/2016MNRAS.456.1172G}
}

@article{Kellermann.etal.1989a,
	author = {{Kellermann}, K.~I. and {Sramek}, R. and {Schmidt}, M. and {Shaffer}, D.~B. and {Green}, R.},
	journal = {\aj},
	month = oct,
	pages = {1195-1207},
	title = {{VLA observations of objects in the Palomar Bright Quasar Survey}},
	volume = 98,
	year = 1989}

@article{KozielWierzbowska.Goyal.Zywucka.2020a,
  author =        {{Kozie{\l}-Wierzbowska}, Dorota and {Goyal}, Arti and
                   {{\.Z}ywucka}, Natalia},
  doi =           {10.3847/1538-4365/ab63d3},
  journal =       {\apjs},
  month =         apr,
  number =        {2},
  pages =         {53},
  title =         {{Radio Sources Associated with Optical Galaxies and
                   Having Unresolved or Extended Morphologies (ROGUE).
                   I. A Catalog of SDSS Galaxies with FIRST Core
                   Identifications}},
  volume =        {247},
  year =          {2020},
  eid =           {53},
  adsurl =        {https://ui.adsabs.harvard.edu/abs/2020ApJS..247...53K},
}

@article{Murphy.etal.2011a,
	author = {{Murphy}, E.~J. and {Condon}, J.~J. and {Schinnerer}, E. and {Kennicutt}, R.~C. and {Calzetti}, D. and {Armus}, L. and {Helou}, G. and {Turner}, J.~L. and {Aniano}, G. and {Beir{\~a}o}, P. and {Bolatto}, A.~D. and {Brandl}, B.~R. and {Croxall}, K.~V. and {Dale}, D.~A. and {Donovan Meyer}, J.~L. and {Draine}, B.~T. and {Engelbracht}, C. and {Hunt}, L.~K. and {Hao}, C.-N. and {Koda}, J. and {Roussel}, H. and {Skibba}, R. and {Smith}, J.-D.~T.},
	journal = {\apj},
	month = aug,
	pages = {67},
	title = {{Calibrating Extinction-free Star Formation Rate Diagnostics with 33 GHz Free-free Emission in NGC 6946}},
	volume = 737,
	year = 2011}

@article{Rusinek.etal.2020a,
	author = {{Rusinek}, Katarzyna and {Sikora}, Marek and {Kozie{\l}-Wierzbowska}, Dorota and {Gupta}, Maitrayee},
	journal = {\apj},
	month = sep,
	number = {2},
	pages = {125},
	title = {{On the Diversity of Jet Production Efficiency in Swift/BAT AGNs}},
	volume = {900},
	year = 2020}

@article{Sikora.Stawarz.Lasota.2007a,
	author = {{Sikora}, M. and {Stawarz}, {\L}. and {Lasota}, J.-P.},
	journal = {\apj},
	month = apr,
	pages = {815-828},
	title = {{Radio Loudness of Active Galactic Nuclei: Observational Facts and Theoretical Implications}},
	volume = 658,
	year = 2007}

@article{Stasinska.etal.2025a,
	author = {{Stasi{\'n}ska}, G. and {Vale Asari}, N. and {W{\'o}jtowicz}, A. and {Kozie{\l}-Wierzbowska}, D.},
	journal = {\aap},
	month = jan,
	pages = {A135},
	title = {{Optically active and optically inactive radio galaxies as sub-populations of the main galaxy sample of the SDSS}},
	volume = {693},
	year = 2025}

@article{Strauss.etal.2002a,
  author =        {{Strauss}, M.~A. and {Weinberg}, D.~H. and
                   {Lupton}, R.~H. and {Narayanan}, V.~K. and
                   {Annis}, J. and {Bernardi}, M. and {Blanton}, M. and
                   {Burles}, S. and {Connolly}, A.~J. and
                   {Dalcanton}, J. and {Doi}, M. and {Eisenstein}, D. and
                   {Frieman}, J.~A. and {Fukugita}, M. and {Gunn}, J.~E. and
                   {Ivezi{\'c}}, {\v Z}. and {Kent}, S. and
                   {Kim}, R.~S.~J. and {Knapp}, G.~R. and {Kron}, R.~G. and
                   {Munn}, J.~A. and {Newberg}, H.~J. and
                   {Nichol}, R.~C. and {Okamura}, S. and {Quinn}, T.~R. and
                   {Richmond}, M.~W. and {Schlegel}, D.~J. and
                   {Shimasaku}, K. and {SubbaRao}, M. and
                   {Szalay}, A.~S. and {Vanden Berk}, D. and
                   {Vogeley}, M.~S. and {Yanny}, B. and {Yasuda}, N. and
                   {York}, D.~G. and {Zehavi}, I.},
  doi =           {10.1086/342343},
  journal =       {\aj},
  month =         sep,
  pages =         {1810-1824},
  title =         {{Spectroscopic Target Selection in the Sloan Digital
                   Sky Survey: The Main Galaxy Sample}},
  volume =        {124},
  year =          {2002},
  url =           {http://dx.doi.org/10.1086/342343},
  adsurl =        {http://adsabs.harvard.edu/abs/2002AJ....124.1810S},
}

@ARTICLE{X12,
       author = {{Xie}, Fu-Guo and {Yuan}, Feng},
        title = "{Radiative efficiency of hot accretion flows}",
      journal = {\mnras},
         year = 2012,
        month = dec,
       volume = {427},
       number = {2},
        pages = {1580-1586},
          doi = {10.1111/j.1365-2966.2012.22030.x},
archivePrefix = {arXiv},
       eprint = {1207.3113},
 primaryClass = {astro-ph.HE},
       adsurl = {https://ui.adsabs.harvard.edu/abs/2012MNRAS.427.1580X}
}

@ARTICLE{X19,
       author = {{Xie}, Fu-Guo and {Zdziarski}, Andrzej A.},
        title = "{Radiative Properties of Magnetically Arrested Disks}",
      journal = {\apj},
         year = 2019,
        month = dec,
       volume = {887},
       number = {2},
          eid = {167},
        pages = {167},
          doi = {10.3847/1538-4357/ab5848},
archivePrefix = {arXiv},
       eprint = {1911.06439},
 primaryClass = {astro-ph.HE},
       adsurl = {https://ui.adsabs.harvard.edu/abs/2019ApJ...887..167X}
}

@ARTICLE{X23,
       author = {{Xie}, Fu-Guo and {Narayan}, Ramesh and {Yuan}, Feng},
        title = "{Observational Constraints on Direct Electron Heating in the Hot Accretion Flows in Sgr A* and M87*}",
      journal = {\apj},
         year = 2023,
        month = jan,
       volume = {942},
       number = {1},
          eid = {20},
        pages = {20},
          doi = {10.3847/1538-4357/aca534},
archivePrefix = {arXiv},
       eprint = {2210.02879},
 primaryClass = {astro-ph.HE},
       adsurl = {https://ui.adsabs.harvard.edu/abs/2023ApJ...942...20X}
}

@ARTICLE{S07,
       author = {{Sharma}, Prateek and {Quataert}, Eliot and {Hammett}, Gregory W. and {Stone}, James M.},
        title = "{Electron Heating in Hot Accretion Flows}",
      journal = {\apj},
         year = 2007,
        month = oct,
       volume = {667},
       number = {2},
        pages = {714-723},
          doi = {10.1086/520800},
archivePrefix = {arXiv},
       eprint = {astro-ph/0703572},
 primaryClass = {astro-ph},
       adsurl = {https://ui.adsabs.harvard.edu/abs/2007ApJ...667..714S}
}

@INPROCEEDINGS{N98,
       author = {{Narayan}, R. and {Mahadevan}, R. and {Quataert}, E.},
        title = "{Advection-dominated accretion around black holes}",
    booktitle = {Theory of Black Hole Accretion Disks},
         year = 1998,
       editor = {{Abramowicz}, M.~A. and {Bj{\"o}rnsson}, G. and {Pringle}, J.~E.},
        month = jan,
        pages = {148-182},
          doi = {10.48550/arXiv.astro-ph/9803141},
archivePrefix = {arXiv},
       eprint = {astro-ph/9803141},
 primaryClass = {astro-ph},
       adsurl = {https://ui.adsabs.harvard.edu/abs/1998tbha.conf..148N}
}

@article{York.etal.2000a,
  author =        {{York}, D.~G. and {Adelman}, J. and
                   {Anderson}, Jr., J.~E. and {Anderson}, S.~F. and
                   {Annis}, J. and {Bahcall}, N.~A. and {Bakken}, J.~A. and
                   {Barkhouser}, R. and {Bastian}, S. and {Berman}, E. and
                   {Boroski}, W.~N. and {Bracker}, S. and {Briegel}, C. and
                   {Briggs}, J.~W. and {Brinkmann}, J. and {Brunner}, R. and
                   {Burles}, S. and {Carey}, L. and {Carr}, M.~A. and
                   {Castander}, F.~J. and {Chen}, B. and
                   {Colestock}, P.~L. and {Connolly}, A.~J. and
                   {Crocker}, J.~H. and {Csabai}, I. and
                   {Czarapata}, P.~C. and {Davis}, J.~E. and {Doi}, M. and
                   {Dombeck}, T. and {Eisenstein}, D. and {Ellman}, N. and
                   {Elms}, B.~R. and {Evans}, M.~L. and {Fan}, X. and
                   {Federwitz}, G.~R. and {Fiscelli}, L. and
                   {Friedman}, S. and {Frieman}, J.~A. and
                   {Fukugita}, M. and {Gillespie}, B. and {Gunn}, J.~E. and
                   {Gurbani}, V.~K. and {de Haas}, E. and {Haldeman}, M. and
                   {Harris}, F.~H. and {Hayes}, J. and {Heckman}, T.~M. and
                   {Hennessy}, G.~S. and {Hindsley}, R.~B. and
                   {Holm}, S. and {Holmgren}, D.~J. and {Huang}, C.-h. and
                   {Hull}, C. and {Husby}, D. and {Ichikawa}, S.-I. and
                   {Ichikawa}, T. and {Ivezi{\'c}}, {\v Z}. and
                   {Kent}, S. and {Kim}, R.~S.~J. and {Kinney}, E. and
                   {Klaene}, M. and {Kleinman}, A.~N. and {Kleinman}, S. and
                   {Knapp}, G.~R. and {Korienek}, J. and {Kron}, R.~G. and
                   {Kunszt}, P.~Z. and {Lamb}, D.~Q. and {Lee}, B. and
                   {Leger}, R.~F. and {Limmongkol}, S. and
                   {Lindenmeyer}, C. and {Long}, D.~C. and {Loomis}, C. and
                   {Loveday}, J. and {Lucinio}, R. and {Lupton}, R.~H. and
                   {MacKinnon}, B. and {Mannery}, E.~J. and
                   {Mantsch}, P.~M. and {Margon}, B. and {McGehee}, P. and
                   {McKay}, T.~A. and {Meiksin}, A. and {Merelli}, A. and
                   {Monet}, D.~G. and {Munn}, J.~A. and
                   {Narayanan}, V.~K. and {Nash}, T. and {Neilsen}, E. and
                   {Neswold}, R. and {Newberg}, H.~J. and
                   {Nichol}, R.~C. and {Nicinski}, T. and {Nonino}, M. and
                   {Okada}, N. and {Okamura}, S. and {Ostriker}, J.~P. and
                   {Owen}, R. and {Pauls}, A.~G. and {Peoples}, J. and
                   {Peterson}, R.~L. and {Petravick}, D. and
                   {Pier}, J.~R. and {Pope}, A. and {Pordes}, R. and
                   {Prosapio}, A. and {Rechenmacher}, R. and
                   {Quinn}, T.~R. and {Richards}, G.~T. and
                   {Richmond}, M.~W. and {Rivetta}, C.~H. and
                   {Rockosi}, C.~M. and {Ruthmansdorfer}, K. and
                   {Sandford}, D. and {Schlegel}, D.~J. and
                   {Schneider}, D.~P. and {Sekiguchi}, M. and
                   {Sergey}, G. and {Shimasaku}, K. and
                   {Siegmund}, W.~A. and {Smee}, S. and {Smith}, J.~A. and
                   {Snedden}, S. and {Stone}, R. and {Stoughton}, C. and
                   {Strauss}, M.~A. and {Stubbs}, C. and {SubbaRao}, M. and
                   {Szalay}, A.~S. and {Szapudi}, I. and
                   {Szokoly}, G.~P. and {Thakar}, A.~R. and
                   {Tremonti}, C. and {Tucker}, D.~L. and {Uomoto}, A. and
                   {Vanden Berk}, D. and {Vogeley}, M.~S. and
                   {Waddell}, P. and {Wang}, S.-i. and {Watanabe}, M. and
                   {Weinberg}, D.~H. and {Yanny}, B. and {Yasuda}, N. and
                   {SDSS Collaboration}},
  doi =           {10.1086/301513},
  journal =       {\aj},
  month =         sep,
  pages =         {1579-1587},
  title =         {{The Sloan Digital Sky Survey: Technical Summary}},
  volume =        {120},
  year =          {2000},
  adsurl =        {http://adsabs.harvard.edu/abs/2000AJ....120.1579Y},
}

@article{Baldwin.Phillips.Terlevich.1981a,
  author =        {{Baldwin}, J.~A. and {Phillips}, M.~M. and
                   {Terlevich}, R.},
  doi =           {10.1086/130766},
  journal =       {\pasp},
  month =         feb,
  pages =         {5-19},
  title =         {{Classification parameters for the emission-line
                   spectra of extragalactic objects}},
  volume =        {93},
  year =          {1981},
  adsurl =        {http://adsabs.harvard.edu/abs/1981PASP...93....5B},
}

@article{Abazajian.etal.2009a,
  author =        {{Abazajian}, K.~N. and {Adelman-McCarthy}, J.~K. and
                   {Ag{\"u}eros}, M.~A. and {Allam}, S.~S. and
                   {Allende Prieto}, C. and {An}, D. and
                   {Anderson}, K.~S.~J. and {Anderson}, S.~F. and
                   {Annis}, J. and {Bahcall}, N.~A. and et al.},
  doi =           {10.1088/0067-0049/182/2/543},
  journal =       {\apjs},
  month =         jun,
  pages =         {543-558},
  title =         {{The Seventh Data Release of the Sloan Digital Sky
                   Survey}},
  volume =        {182},
  year =          {2009},
  eid =           {543},
  url =           {http://dx.doi.org/10.1088/0067-0049/182/2/543},
  adsurl =        {http://adsabs.harvard.edu/abs/2009ApJS..182..543A},
}

@article{White.etal.1997a,
  author =        {{White}, Richard L. and {Becker}, Robert H. and
                   {Helfand}, David J. and {Gregg}, Michael D.},
  doi =           {10.1086/303564},
  journal =       {\apj},
  month =         feb,
  number =        {2},
  pages =         {479-493},
  title =         {{A Catalog of 1.4 GHz Radio Sources from the FIRST
                   Survey}},
  volume =        {475},
  year =          {1997},
  adsurl =        {https://ui.adsabs.harvard.edu/abs/1997ApJ...475..479W},
}

@article{Condon.etal.1998a,
  author =        {{Condon}, J.~J. and {Cotton}, W.~D. and
                   {Greisen}, E.~W. and {Yin}, Q.~F. and {Perley}, R.~A. and
                   {Taylor}, G.~B. and {Broderick}, J.~J.},
  doi =           {10.1086/300337},
  journal =       {\aj},
  month =         may,
  number =        {5},
  pages =         {1693-1716},
  title =         {{The NRAO VLA Sky Survey}},
  volume =        {115},
  year =          {1998},
  adsurl =        {https://ui.adsabs.harvard.edu/abs/1998AJ....115.1693C},
}

@article{CidFernandes.etal.2005a,
  author =        {{Cid Fernandes}, R. and {Mateus}, A. and
                   {Sodr{\'e}}, L. and {Stasi{\'n}ska}, G. and
                   {Gomes}, J.~M.},
  doi =           {10.1111/j.1365-2966.2005.08752.x},
  journal =       {\mnras},
  month =         apr,
  pages =         {363-378},
  title =         {{Semi-empirical analysis of Sloan Digital Sky Survey
                   galaxies - I. Spectral synthesis method}},
  volume =        {358},
  year =          {2005},
  adsurl =        {http://adsabs.harvard.edu/abs/2005MNRAS.358..363C},
}

@article{Tremaine.etal.2002a,
  author =        {{Tremaine}, S. and {Gebhardt}, K. and {Bender}, R. and
                   {Bower}, G. and {Dressler}, A. and {Faber}, S.~M. and
                   {Filippenko}, A.~V. and {Green}, R. and
                   {Grillmair}, C. and {Ho}, L.~C. and {Kormendy}, J. and
                   {Lauer}, T.~R. and {Magorrian}, J. and {Pinkney}, J. and
                   {Richstone}, D.},
  doi =           {10.1086/341002},
  journal =       {\apj},
  month =         aug,
  pages =         {740-753},
  title =         {{The Slope of the Black Hole Mass versus Velocity
                   Dispersion Correlation}},
  volume =        {574},
  year =          {2002},
  adsurl =        {http://adsabs.harvard.edu/abs/2002ApJ...574..740T},
}

@article{Stasinska.etal.2006a,
  author =        {{Stasi{\'n}ska}, G. and {Cid Fernandes}, R. and
                   {Mateus}, A. and {Sodr{\'e}}, L. and {Asari}, N.~V.},
  doi =           {10.1111/j.1365-2966.2006.10732.x},
  journal =       {\mnras},
  month =         sep,
  pages =         {972-982},
  title =         {{Semi-empirical analysis of Sloan Digital Sky Survey
                   galaxies - III. How to distinguish AGN hosts}},
  volume =        {371},
  year =          {2006},
  adsurl =        {http://adsabs.harvard.edu/abs/2006MNRAS.371..972S},
}

@article{KozielWierzbowska.etal.2021a,
  author =        {{Kozie{\l}-Wierzbowska}, D. and {Vale Asari}, N. and
                   {Stasi{\'n}ska}, G. and {Herpich}, F.~R. and
                   {Sikora}, M. and {{\.Z}ywucka}, N. and {Goyal}, A.},
  doi =           {10.3847/1538-4357/abe308},
  journal =       {\apj},
  month =         mar,
  number =        {1},
  pages =         {64},
  title =         {{Identifying Radio-active Galactic Nuclei among
                   Radio-emitting Galaxies}},
  volume =        {910},
  year =          {2021},
  eid =           {64},
  adsurl =        {https://ui.adsabs.harvard.edu/abs/2021ApJ...910...64K},
}

@article{Ekholm.etal.2001a,
  author =        {{Ekholm}, T. and {Baryshev}, Y. and {Teerikorpi}, P. and
                   {Hanski}, M.~O. and {Paturel}, G.},
  doi =           {10.1051/0004-6361:20010161},
  journal =       {\aap},
  month =         mar,
  pages =         {L17-L20},
  title =         {{On the quiescence of the Hubble flow in the vicinity
                   of the Local Group. A study using galaxies with
                   distances from the Cepheid PL-relation}},
  volume =        {368},
  year =          {2001},
  url =           {http://dx.doi.org/10.1051/0004-6361:20010161},
  adsurl =        {http://adsabs.harvard.edu/abs/2001A%26A...368L..17E},
}

@article{CidFernandes.etal.2011a,
  author =        {{Cid Fernandes}, R. and {Stasi{\'n}ska}, G. and
                   {Mateus}, A. and {Vale Asari}, N.},
  doi =           {10.1111/j.1365-2966.2011.18244.x},
  journal =       {\mnras},
  month =         may,
  pages =         {1687-1699},
  title =         {{A comprehensive classification of galaxies in the
                   Sloan Digital Sky Survey: how to tell true from fake
                   AGN?}},
  volume =        {413},
  year =          {2011},
  adsurl =        {http://adsabs.harvard.edu/abs/2011MNRAS.413.1687C},
}

@article{Stasinska.etal.2008a,
  author =        {{Stasi{\'n}ska}, G. and {Vale Asari}, N. and
                   {Cid Fernandes}, R. and {Gomes}, J.~M. and
                   {Schlickmann}, M. and {Mateus}, A. and
                   {Schoenell}, W. and {Sodr{\'e}}, Jr., L.},
  doi =           {10.1111/j.1745-3933.2008.00550.x},
  journal =       {\mnras},
  month =         nov,
  pages =         {L29-L33},
  title =         {{Can retired galaxies mimic active galaxies? Clues
                   from the Sloan Digital Sky Survey}},
  volume =        {391},
  year =          {2008},
  adsurl =        {http://adsabs.harvard.edu/abs/2008MNRAS.391L..29S},
}

@article{Spinoglio.FernandezOntiveros.Malkan.2024a,
  author =        {{Spinoglio}, Luigi and
                   {Fern{\'a}ndez-Ontiveros}, Juan Antonio and
                   {Malkan}, Matthew A.},
  doi =           {10.3847/1538-4357/ad23e4},
  journal =       {\apj},
  month =         apr,
  number =        {2},
  pages =         {117},
  title =         {{The Spectral Energy Distributions and Bolometric
                   Luminosities of Local AGN: Study of the Complete 12
                   {\ensuremath{\mu}}m AGN Sample}},
  volume =        {964},
  year =          {2024},
  eid =           {117},
  adsurl =        {https://ui.adsabs.harvard.edu/abs/2024ApJ...964..117S},
}

@article{Jin.Ward.Done.2012a,
  author =        {{Jin}, C. and {Ward}, M. and {Done}, C.},
  doi =           {10.1111/j.1365-2966.2012.21272.x},
  journal =       {\mnras},
  month =         sep,
  pages =         {907-929},
  title =         {{A combined optical and X-ray study of unobscured
                   type 1 active galactic nuclei - III. Broad-band SED
                   properties}},
  volume =        {425},
  year =          {2012},
  adsurl =        {http://adsabs.harvard.edu/abs/2012MNRAS.425..907J},
}

@article{Jin.etal.2018b,
  author =        {{Jin}, J.-J. and {Zhu}, Y.-N. and {Meng}, X.-M. and
                   {Lei}, F.-J. and {Wu}, H.},
  doi =           {10.3847/1538-4357/aad4f7},
  journal =       {\apj},
  month =         sep,
  pages =         {32},
  title =         {{A Systematic Analysis of Stellar Populations in the
                   Host Galaxies of SDSS Type I QSOs}},
  volume =        {864},
  year =          {2018},
  eid =           {32},
  adsurl =        {http://adsabs.harvard.edu/abs/2018ApJ...864...32J},
}

@ARTICLE{Wojtowicz2023,
       author = {{W{\'o}jtowicz}, Anna and {Stawarz}, {\L}ukasz and {Cheung}, C.~C. and {Werner}, Norbert and {Rudka}, Dominik},
        title = "{Radio Emission of Nearby Early-type Galaxies in the Low and Very Low Radio Luminosity Range}",
      journal = {\apj},
         year = 2023,
        month = feb,
       volume = {944},
       number = {2},
          eid = {195},
        pages = {195},
          doi = {10.3847/1538-4357/acb498},
archivePrefix = {arXiv},
       eprint = {2209.14638},
 primaryClass = {astro-ph.GA},
       adsurl = {https://ui.adsabs.harvard.edu/abs/2023ApJ...944..195W}
}

@ARTICLE{Capetti2022,
       author = {{Capetti}, A. and {Brienza}, M. and {Balmaverde}, B. and {Best}, P.~N. and {Baldi}, R.~D. and {Drabent}, A. and {G{\"u}rkan}, G. and {Rottgering}, H.~J.~A. and {Tasse}, C. and {Webster}, B.},
        title = "{The LOFAR view of giant, early-type galaxies: Radio emission from active nuclei and star formation}",
      journal = {\aap},
         year = 2022,
        month = apr,
       volume = {660},
          eid = {A93},
        pages = {A93},
          doi = {10.1051/0004-6361/202142911},
archivePrefix = {arXiv},
       eprint = {2202.08593},
 primaryClass = {astro-ph.GA},
       adsurl = {https://ui.adsabs.harvard.edu/abs/2022A&A...660A..93C}
}

@ARTICLE{Padovani.etal.2015a,
       author = {{Padovani}, P. and {Bonzini}, M. and {Kellermann}, K.~I. and {Miller}, N. and {Mainieri}, V. and {Tozzi}, P.},
        title = "{Radio-faint AGN: a tale of two populations}",
      journal = {\mnras},
         year = 2015,
        month = sep,
       volume = {452},
       number = {2},
        pages = {1263-1279},
          doi = {10.1093/mnras/stv1375},
archivePrefix = {arXiv},
       eprint = {1506.06554},
 primaryClass = {astro-ph.GA},
       adsurl = {https://ui.adsabs.harvard.edu/abs/2015MNRAS.452.1263P}
}

@article{Igumenshchev.Narayan.Abramowicz.2003a,
  author =        {{Igumenshchev}, Igor V. and {Narayan}, Ramesh and
                   {Abramowicz}, Marek A.},
  doi =           {10.1086/375769},
  journal =       {\apj},
  month =         aug,
  number =        {2},
  pages =         {1042-1059},
  title =         {{Three-dimensional Magnetohydrodynamic Simulations of
                   Radiatively Inefficient Accretion Flows}},
  volume =        {592},
  year =          {2003},
  adsurl =        {https://ui.adsabs.harvard.edu/abs/2003ApJ...592.1042I},
}

@ARTICLE{Beckmann2024,
       author = {{Beckmann}, R.~S. and {Smethurst}, R.~J. and {Simmons}, B.~D. and {Coil}, A. and {Dubois}, Y. and {Garland}, I.~L. and {Lintott}, C.~J. and {Martin}, G. and {Peirani}, S. and {Pichon}, C.},
        title = "{Supermassive black holes in merger-free galaxies have higher spins which are preferentially aligned with their host galaxy}",
      journal = {\mnras},
         year = 2024,
        month = feb,
       volume = {527},
       number = {4},
        pages = {10867-10877},
          doi = {10.1093/mnras/stad1795},
       adsurl = {https://ui.adsabs.harvard.edu/abs/2024MNRAS.52710867B}
}

@article{Blandford.Znajek.1977a,
  author =        {{Blandford}, R.~D. and {Znajek}, R.~L.},
  doi =           {10.1093/mnras/179.3.433},
  journal =       {\mnras},
  month =         may,
  pages =         {433-456},
  title =         {{Electromagnetic extraction of energy from Kerr black
                   holes.}},
  volume =        {179},
  year =          {1977},
  adsurl =        {https://ui.adsabs.harvard.edu/abs/1977MNRAS.179..433B},
}

@article{Tchekhovskoy.Narayan.McKinney.2011a,
  author =        {{Tchekhovskoy}, Alexander and {Narayan}, Ramesh and
                   {McKinney}, Jonathan C.},
  doi =           {10.1111/j.1745-3933.2011.01147.x},
  journal =       {\mnras},
  month =         nov,
  number =        {1},
  pages =         {L79-L83},
  title =         {{Efficient generation of jets from magnetically
                   arrested accretion on a rapidly spinning black hole}},
  volume =        {418},
  year =          {2011},
  adsurl =        {https://ui.adsabs.harvard.edu/abs/2011MNRAS.418L..79T},
}

@ARTICLE{Miller.etal.1990a,
       author = {{Miller}, L. and {Peacock}, J.~A. and {Mead}, A.~R.~G.},
        title = "{The bimodal radio luminosity function of quasars.}",
      journal = {\mnras},
         year = 1990,
        month = may,
       volume = {244},
        pages = {207-213},
       adsurl = {https://ui.adsabs.harvard.edu/abs/1990MNRAS.244..207M}
}

@ARTICLE{White.etal.2000a,
       author = {{White}, Richard L. and {Becker}, Robert H. and {Gregg}, Michael D. and {Laurent-Muehleisen}, Sally A. and {Brotherton}, Michael S. and {Impey}, Chris D. and {Petry}, Catherine E. and {Foltz}, Craig B. and {Chaffee}, Frederic H. and {Richards}, Gordon T. and {Oegerle}, William R. and {Helfand}, David J. and {McMahon}, Richard G. and {Cabanela}, Juan E.},
        title = "{The FIRST Bright Quasar Survey. II. 60 Nights and 1200 Spectra Later}",
      journal = {\apjs},
         year = 2000,
        month = feb,
       volume = {126},
       number = {2},
        pages = {133-207},
          doi = {10.1086/313300},
archivePrefix = {arXiv},
       eprint = {astro-ph/9912215},
 primaryClass = {astro-ph},
       adsurl = {https://ui.adsabs.harvard.edu/abs/2000ApJS..126..133W}
}

@ARTICLE{Rafter.etal.2009a,
       author = {{Rafter}, Stephen E. and {Crenshaw}, D. Michael and {Wiita}, Paul J.},
        title = "{Radio Properties of Low-Redshift Broad Line Active Galactic Nuclei}",
      journal = {\aj},
         year = 2009,
        month = jan,
       volume = {137},
       number = {1},
        pages = {42-52},
          doi = {10.1088/0004-6256/137/1/42},
archivePrefix = {arXiv},
       eprint = {0809.3977},
 primaryClass = {astro-ph},
       adsurl = {https://ui.adsabs.harvard.edu/abs/2009AJ....137...42R}
}

@ARTICLE{Singal.etal.2013a,
       author = {{Singal}, J. and {Petrosian}, V. and {Stawarz}, {\L}. and {Lawrence}, A.},
        title = "{The Radio and Optical Luminosity Evolution of Quasars. II. The SDSS Sample}",
      journal = {\apj},
         year = 2013,
        month = feb,
       volume = {764},
       number = {1},
          eid = {43},
        pages = {43},
          doi = {10.1088/0004-637X/764/1/43},
archivePrefix = {arXiv},
       eprint = {1207.3396},
 primaryClass = {astro-ph.CO},
       adsurl = {https://ui.adsabs.harvard.edu/abs/2013ApJ...764...43S}
}

@ARTICLE{White.etal.2007a,
       author = {{White}, Richard L. and {Helfand}, David J. and {Becker}, Robert H. and {Glikman}, Eilat and {de Vries}, Wim},
        title = "{Signals from the Noise: Image Stacking for Quasars in the FIRST Survey}",
      journal = {\apj},
         year = 2007,
        month = jan,
       volume = {654},
       number = {1},
        pages = {99-114},
          doi = {10.1086/507700},
archivePrefix = {arXiv},
       eprint = {astro-ph/0607335},
 primaryClass = {astro-ph},
       adsurl = {https://ui.adsabs.harvard.edu/abs/2007ApJ...654...99W}
}

@ARTICLE{Bicknell.2002a,
       author = {{Bicknell}, Geoffrey V.},
        title = "{Connections between jet physics and the properties of radio-loud and radio-quiet galaxies}",
      journal = {\nar},
         year = 2002,
        month = may,
       volume = {46},
       number = {2-7},
        pages = {365-379},
          doi = {10.1016/S1387-6473(01)00210-X},
       adsurl = {https://ui.adsabs.harvard.edu/abs/2002NewAR..46..365B}
}

\appendix
\section{Blandford-Znajek jet power}
\label{sec:appendixA}

The total power of a Blandford-Znajek (BZ) jet, as formulated in \citet{Tchekhovskoy.Narayan.McKinney.2011a}, is
\begin{equation}
    P_j \simeq \frac{\kappa}{4 \pi} \, \Omega^2 \, \Phi_{{\rm BH}}^2 \, \tilde{f}\!(\Omega) \, ,
\end{equation}
where $\kappa \simeq 0.05$ is a numerical constant, the angular frequency of the BH horizon is $\Omega = a \, c / 2 \, R_{\rm H}$, the dimensionless BH spin parameter is $a \in [0,1]$, the BH event horizon radius is $R_{\rm H} = R_g \, (1 + \sqrt{1-a^2})$, the gravitational radius of the BH with mass $M_{{\rm BH}}$ is $R_g = G M_{{\rm BH}}/c^2$, the magnetic flux threading the BH horizon is denoted by $\Phi_{{\rm BH}}$, and finally the function $\tilde{f}\!(\Omega) \simeq 1 + 1.38 \, (\Omega \, R_g/c)^2 - 9.2 (\Omega \, R_g/c)^4$.

% By introducing the saturation value of the magnetic flux, corresponding
The saturation value of the magnetic flux corresponding to the `magnetically arrested disk' (MAD) regime \citep{Igumenshchev.Narayan.Abramowicz.2003a} is
\begin{equation}
    \Phi_{\rm MAD} = \eta \, \sqrt{\dot{M}_{\rm acc} c \, R_g^2} \, ,
\end{equation}
where the numerical constant $\eta \simeq 50$ and  $\dot{M}_{\rm acc}$ is the mass accretion rate, one may define the dimensionless magnetic flux parameter
\begin{equation}
    \varphi \equiv \frac{\Phi_{{\rm BH}}}{\Phi_{\rm MAD}}  \, \leq 1 \, .
\end{equation}
%With such, one
From these definitions, one may re-write the BZ total jet power as
\begin{equation}
    P_j \simeq F\!(a) \,\, \varphi^2 \,\, \dot{M}_{\rm acc} c^2 \, ,
\end{equation}
where 
\begin{eqnarray}
    F\!(a) & = & \frac{\kappa \, \eta}{4 \pi} \,\, x^2  f(x) \, , \nonumber \\ 
     x &= & \frac{a}{2 \, (1+\sqrt{1-a^2})} \, , \quad \textrm{and} \nonumber \\
    f(x) &=& 1 + 1.38 x^2 - 9.2 x^4 \, .
\end{eqnarray}
As shown in Figure\,\ref{fig:spin}, for moderate spin values, $F(a=0.5) \simeq 0.2$, increasing approximately as $a^2$ by one order of magnitude up to $F(a=1) \simeq 2$.

\begin{figure}[th]
\centering
\includegraphics[width=\columnwidth]{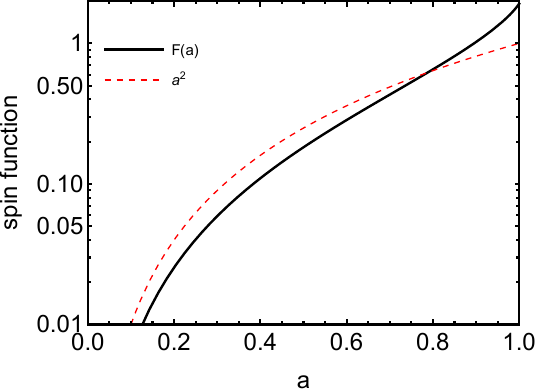}
\caption{Spin dependence of the Blandford-Znajek jet.} 
\label{fig:spin}
\end{figure}

\section{Accretion radiative efficiency}
\label{sec:appendixB}

The radiative efficiency of accretion disks and flows is defined as
\begin{equation}
    \epsilon \equiv \frac{L_{\rm bol}}{\dot{M}_{\rm acc} c^2} \, ,
\end{equation}
where $\dot{M}_{\rm acc}$ is the mass accretion rate and $L_{\rm bol}$ is the bolometric accretion disk luminosity. The consensus is that, while at high accretion rates the disk radiative efficiency remains approximately constant at a value $\epsilon_0$ of a few to a few tens of percent (depending only weakly on the black hole spin), at lower accretion rates, \ie within the regime of `radiatively inefficient accretion flows', the efficiency decreases approximately linearly with the Eddington-normalized mass accretion rate \citep[see, e.g.,][]{N98,S07,X12,X19,X23}. We can therefore parametrize 
\begin{equation}
\epsilon \simeq \left\{ 
  \begin{array}{ c l }
   \epsilon_0 \, \frac{\dot{m}}{\dot{m}_{\rm cr}}  & \quad \textrm{if } \quad \dot{m} < \dot{m}_{\rm cr}, \\
   \epsilon_0  & \quad \textrm{if } \quad \dot{m} \geq \dot{m}_{\rm cr}, \\
  \end{array}
\right.   \label{eq:r}
\end{equation}
where 
\begin{eqnarray}
      \dot{m} &\equiv &\frac{\dot{M}_{\rm acc}}{\dot{M}_{\rm Edd}}, \, \nonumber \\ \dot{M}_{\rm Edd} c^2 & = & \frac{L_{\rm Edd}}{\epsilon_0} \,  , \quad \textrm{and} \nonumber \\
    L_{\rm Edd} &=& 4 \pi G M_{{\rm BH}} m_p c / \sigma_{\rm T} \, .
\end{eqnarray}
What remains most uncertain in this context is the exact value of the critical accretion rate $\dot{m}_{\rm cr}$. In particular, while canonical general considerations suggest $\dot{m}_{\rm cr} \sim 10^{-2}$ \citep{N98}, some numerical simulations \citep{S07}, as well as the most recent calculations carried out in the MAD limit for the nearby supermassive black holes M87 and Sgr A* \citep{X23}, suggest that accretion disks may remain relatively efficient at the few-percent level down to much lower critical mass accretion rates than previously thought.

Note that, with the above scaling of the disk radiative efficiency $\epsilon$, the accretion Eddington parameter is then expected to scale as 
\begin{equation}
    \lambda \equiv \frac{L_{\rm bol}}{L_{\rm Edd}} = \frac{\epsilon}{\epsilon_0} \, \dot{m} \,\, \simeq \left\{ 
  \begin{array}{ c l }
   \dot{m}^2/\dot{m}_{\rm cr}  & \quad \textrm{if } \quad \lambda < \dot{m}_{\rm cr}, \\
   \dot{m}  & \quad \textrm{if } \quad \lambda \geq \dot{m}_{\rm cr}. \\
  \end{array}
\right.
\end{equation}

\end{document}